\documentclass[aps,prd,twocolumn,superscriptaddress,showpacs]{revtex4-1}
\usepackage[T1]{fontenc} 
\usepackage{mathptmx} 
\usepackage{tabularx} 
\usepackage{amsmath,amssymb,tensor,physics,overarrows,empheq}  
\usepackage{babel,csquotes,xpatch}

\usepackage{amssymb,latexsym}
\usepackage{amsmath}
\usepackage{tensor}
\usepackage[utf8]{inputenc}
\usepackage[colorlinks=true,linkcolor=blue,citecolor=blue,urlcolor=blue]{hyperref}
\usepackage{graphicx}     
\usepackage[dvipsnames]{xcolor}
\usepackage{floatflt}     
\usepackage{wrapfig}
\usepackage{color}
\usepackage{pstricks}
\usepackage{subcaption}
\usepackage{color}
\definecolor{DarkRed}{rgb}{0.35,0.01,0.01}
\colorlet{RED}{red}

 \definecolor{Linen}{rgb}{0.98,0.98,0.94}
 \definecolor{Blue}{rgb}{0.,0.,1.0}
 \definecolor{DarkBlue}{rgb}{0.099,0.099,0.44}
 \definecolor{DarkGreen}{rgb}{0.0,0.4,0.0}
 \definecolor{Turquoise}{rgb}{0.0,0.9,0.7}
 \usepackage{tikz}
\usepackage{tkz-euclide}
\usepackage{tikz-3dplot}
\usepackage{pgfplots,pgfplotstable}
\usepackage[makeroom]{cancel}
\pgfplotsset{compat=newest}
\usetikzlibrary{shapes,shapes.arrows,shapes.symbols,shapes.geometric, calc,positioning}
\usepackage{mathtools}
\tikzstyle{start} = [rectangle, rounded corners, 
minimum width=3cm, 
minimum height=0.7cm,
text centered, 
text width=3cm, 
draw=black, 
fill=orange!30]

\tikzstyle{goal2} = [rectangle, rounded corners, 
minimum width=3cm, 
minimum height=0.7cm,
text centered, 
text width=3cm, 
draw=black, 
fill=red!30]

\tikzstyle{goal1} = [rectangle, rounded corners, 
minimum width=3cm, 
minimum height=0.7cm,
text centered, 
text width=3cm, 
draw=black, 
fill=purple!30]

\tikzstyle{goal3} = [rectangle, rounded corners, 
minimum width=3cm, 
minimum height=0.7cm,
text centered, 
text width=2.5cm, 
draw=black, 
fill=green!30]
\tikzstyle{arrow} = [thick,->,>=stealth] 
\begin{document}

 \title{Plasma effects on gravitational lensing and shadow observables of a Kerr-like black hole in a dark matter halo}
\author{Connor McMillin}
\email[]{mcmillin@grinnell.edu}
\affiliation{Department of Physics, Grinnell College, Grinnell, IA 50112}

\author{Zhichen Guan}
\email[]{guanzhic@grinnell.edu}
\affiliation{Department of Physics, Grinnell College, Grinnell, IA 50112}

\author{Owen Gartlan}
\email[]{gartlano@grinnell.edu}
\affiliation{Department of Physics, Grinnell College, Grinnell, IA 50112}

\author{Lotus Liu}
\email[]{lotusliu@uchicago.edu}
\affiliation{Department of Physics, University of Chicago, Chicago, IL 60637}

\author{Leo Rodriguez}
\email[]{rodriguezl@grinnell.edu}
\affiliation{Department of Physics, Grinnell College, Grinnell, IA 50112}
\affiliation{Department of Physics, Worcester Polytechnic Institute, Worcester, MA 01609}  
\author{Shanshan Rodriguez}
\email{rodriguezs@grinnell.edu}
\thanks{Corresponding author.}
\affiliation{Department of Physics, Grinnell College, Grinnell, IA 50112}
\affiliation{Department of Physics, Worcester Polytechnic Institute, Worcester, MA 01609}
\affiliation{Center for Astrophysics, Harvard \& Smithsonian, Cambridge, MA 02138}


\date{\today}
\begin{abstract}

Plasma surrounding a black hole modifies light propagation and can alter the observed shadow, potentially affecting the interpretation of Event Horizon Telescope data. We study the effects of dark matter and nonmagnetized pressureless plasma on the shadow of a Kerr-like black hole by analyzing null geodesics in both homogeneous and inhomogeneous plasma distributions. For the homogeneous plasma profile, the asymptotic Bardeen coordinates acquire a refractive normalization factor arising from the leading-order coupling between the metric function $\Delta(r)$ and the radial plasma function $f_r(r)\propto r^2$. We show that increasing the black hole spin generally enlarges the shadow radius and increases its deformation, while moving the observer away from the equatorial plane decreases both quantities. For the parameter ranges considered, astrophysically reasonable dark matter densities in this model do not produce appreciable changes in the photon trajectories. Plasma effects, however, are significant: increasing the plasma density increases the shadow radius and deformation for homogeneous plasma, but decreases them for inhomogeneous plasma. The energy emission rate likewise depends strongly on the plasma model, with homogeneous plasma producing a substantially larger rate as the plasma strength increases. As an illustrative benchmark, we compare the resulting geometric critical-curve sizes with EHT-inferred shadow-size intervals for M87* and Sgr A*.

\end{abstract}
\maketitle
\section{INTRODUCTION}

Black-hole shadows and strong-field gravitational lensing have become powerful observational tools for probing gravity and the near-horizon environment of compact objects. The first horizon-scale image of the supermassive black hole in M87* released by the Event Horizon Telescope (EHT) marked a major milestone, revealing a bright, asymmetric ring surrounding a central brightness depression whose angular scale is consistent with the characteristic photon-orbit size predicted by general relativity \cite{EHT_first_2019}. This result demonstrated that shadow observables can be measured with sufficient precision to test theoretical models. The subsequent EHT image of Sgr A* extended these tests to our Galactic center, despite the additional challenges posed by rapid source variability and scattering effects \cite{EventHorizonTelescope:2022urf}. These observations motivate continued theoretical efforts to understand how both spacetime geometry and astrophysical environments influence gravitational lensing and the apparent size and shape of black-hole shadows \cite{Synge1966, 1973blho.conf..215B, PerlickTsupko2015}. In realistic environments, however, light does not propagate in a vacuum: plasma around the black hole introduces refraction and dispersion, so the apparent shadow and lensing observables become frequency dependent \cite{Perlick2000RayOF}. A widely used starting point is the Hamiltonian/refractive-index formulation for photon propagation in plasma on curved spacetimes, developed in different forms for magnetized and non-magnetized media \cite{PhysRevD.87.124009}. This framework underlies much of the modern literature connecting theoretical shadow/lensing predictions to astrophysical settings and to EHT-scale observations.

A substantial body of work has explored what plasma does to shadow size and deformation under a variety of models. For static, spherically symmetric spacetimes, Perlick, Tsupko, and Bisnovatyi-Kogan derived analytic expressions for the angular shadow size in a non-magnetized, pressureless plasma with general radial density profiles, and showed explicitly how the shadow radius becomes frequency dependent; they also discussed accreting (infalling) plasma examples relevant for astrophysical applications \cite{PerlickTsupko2015, PerlickTsupkoBisnovatyiKogan2015}. For rotating spacetimes, progress accelerated once separability conditions were identified: Perlick and Tsupko showed that certain classes of plasma frequency profiles preserve separability of the Hamilton-Jacobi equation in Kerr spacetime, enabling semi-analytic characterization of photon regions and shadow boundaries \cite{PerlickTsupko2017KerrPlasma}. The most commonly explored non-magnetized, pressureless plasma models include homogeneous plasmas and inhomogeneous power-law or radially varying density profiles \cite{Bisnovatyi-Kogan:2008qbk, Bisnovatyi-Kogan:2010flt, ErMao2014, Rogers:2015dla, Morozova:2013uyv, Atamurotov:2015nra, PhysRevD.104.064039, Yan:2019etp}. More recent studies have considered more structured or anisotropic plasma distributions, such as Gaussian-type profiles, as well as the effects of magnetized plasma, further emphasizing that plasma modeling is essential for a realistic interpretation of shadow and lensing observables \cite{Zhang:2022osx, Khodadi:2022ulo, Pahlavon:2024caj, Gohain:2025zau}. These results establish a fairly robust qualitative picture: plasma generically induces frequency-dependent shifts in the effective photon region, changing the shadow’s “areal” size and its deformation measures, with the sign and magnitude controlled by how plasma is distributed (uniform vs. centrally concentrated vs. anisotropic), and by spin/inclination. The influence of plasma on black hole shadows has also been extensively studied in general relativity and modified gravity theories \cite{PhysRevD.87.124009, Bisnovatyi-Kogan:2010flt, Er:2013efa, Bisnovatyi-Kogan:2017kii, Liu:2016eju, Crisnejo:2018uyn, Crisnejo:2019ril, Crisnejo:2018ppm, Kumaran:2021rgj, Atamurotov:2021qds, Sun:2022ujt, Bisnovatyi-Kogan:2022yzj, Feleppa:2024vdk, Abdujabbarov:2015pqp, Huang:2018rfn, Atamurotov:2021cgh, Chowdhuri:2020ipb, Ozel:2021ayr, Das:2021otl, Badia:2022phg, Das:2022tqi, Fathi:2023xeg, Raza:2024zkp, Ali:2024cti, Ali:2025dje, Yasmin:2025rnj, Badia:2021kpk, EventHorizonTelescope:2021btj, Babar:2020txt, Kala:2025fld, Kumar:2024vdh,  Feng:2024iqj,  KumarSahoo:2025igt, banerjeeSilhouetteM87New2020, chowdhuryAccretingSchwarzschildlikeCompact2024}. Comprehensive reviews have summarized these developments and clarified the regimes in which plasma effects are expected to be most relevant \cite{PerlickReviewShadows2021}.
 
In addition, the nature of dark matter, which constitutes the dominant mass component of galaxies and ~27\% of the cosmic energy budget, remains a central open problem. While general relativity accurately describes its gravitational effects, it provides no insight into its microscopic origin or particle content. Astrophysical evidence, from galactic rotation curves to colliding clusters, indicates that dark matter forms extended halos in which the most massive galaxies host a central supermassive black hole \cite{Rubin:1980zd, Persic:1995ru, Bertone:2004pz, Clowe:2006eq, Bertone:2016nfn}. Consequently, astrophysical black holes are not isolated systems but are embedded in matter environments. Since dark matter exhibits a weak standard model interaction, an effective way to study its astrophysical properties is to probe its gravitational influence by modifying strong-field dynamics such as particle geodesics, gravitational lensing, and shadow deformation. In particular, dark matter distributions around supermassive black holes may affect accretion, feedback, and gravitational-wave sources such as extreme and intermediate mass-ratio inspirals \cite{PhysRevLett.83.1719, PhysRevD.102.103022, PhysRevD.77.064023}. Studying black holes in realistic matter backgrounds, therefore, provides a promising avenue to probe dark matter properties and test gravity in complex astrophysical settings.

Analytical studies commonly model dark matter halos using parametric density profiles motivated by simulations and galactic dynamics, such as the Navarro-Frenk-White (NFW) profile \cite{Navarro1996,Navarro1997}, Dehnen-type profiles \cite{Dehnen:1993uh, Gohain:2024eer, Al-Badawi:2024asn}, the Burkert model \cite{Burkert1995}, and the Einasto profile, which offers greater flexibility in describing the inner halo structure and is often favored in high-resolution simulations \cite{Einasto1965,BaesEinasto2022}. Simpler analytic profiles, such as the Hernquist model, are also frequently used as tractable approximations to galactic potentials \cite{Hernquist1990, Dai:2023cft}. Recent work has investigated how these dark-matter distributions modify gravitational lensing and shadow properties, either through effective spacetime models or perturbative treatments, and has shown that halo parameters can lead to measurable shifts in shadow size and related observables \cite{Jusufi:2019nrn, Liu:2023oab, Al-Badawi:2025njy, Liang:2025vux, Uktamov:2025lsq}, especially when combined with other environmental effects such as plasma \cite{PhysRevD.107.064040}. These studies help disentangle how much of an observed size or deformation shift arises from dispersive photon propagation versus genuine modifications of the effective gravitational potential.

Motivated by these developments, this paper examines the combined influence of plasma and a dark-matter halo on the gravitational lensing and shadow of a Kerr-like black hole. We consider both homogeneous and inhomogeneous plasma models, adopt an Einasto profile for the dark-matter halo, and analyze how plasma frequency and halo parameters affect shadow size, deformation, and the associated energy emission rate, with an illustrative geometric comparison to EHT-inferred shadow-size intervals for M87* and Sgr A* \cite{EHT_first_2019,EventHorizonTelescope:2022urf}.

The paper is organized as follows. In Section~\ref{sect:DMHaloMetric}, we briefly introduce the Kerr-like black hole in a dark matter halo and its horizon structure. In Section~\ref{Sect:PlasmaInclusionDeriv}, we introduce the plasma models used in geometrical optics around the black hole concerned. In Sections~\ref{sect:GeoShadowDeriv}, \ref{sect:ResultsAnalysis}, and \ref{sect:EERate}, we analyze the plasma influence on photon orbits, deflection angle, black hole shadow and observables, and energy emission rate. In Section \ref{sect:CEHTM87As}, we make an illustrative geometric comparison between our critical-curve sizes and EHT-inferred intervals for M87* and Sgr A*, and report which plasma-frequency ranges remain compatible under this idealized mapping. We summarize our results in Section~\ref{sect:ConclFW}. We have assumed geometrized units $\hslash=c = G = 1$ throughout this work.

\section{Kerr-Like Black Hole in a Dark Matter Halo} \label{sect:DMHaloMetric}

For decades, computational studies have sought to model the internal structure of dark matter halos and understand how their density varies with radius. Early simulations often employed simplified analytic forms, such as the NFW profile, to describe the radial dependence of halo density \cite{Navarro1996, Navarro1997}. More recent high-resolution simulations, however, have revealed systematic deviations from these earlier models, particularly in the innermost and outermost regions of halos \cite{Wang:2019ftp}. As a more flexible and accurate alternative, the Einasto profile is expressed as an exponential function of a power law, which allows the logarithmic slope of the halo density to vary smoothly with radius as approximately $\dv{\ln{\rho}}{\ln r} \propto r^{\alpha}$. Owing to this adaptability, the Einasto profile has been shown to reproduce the radial density distributions of simulated dark matter halos across a wide range of masses and cosmological conditions with remarkable precision \cite{wang_universal_2020}. Specifically, it requires that \cite{Einasto:1965czb, BaesEinasto2022, Liu:2023oab}
\begin{equation}
\rho(r) = \rho_e \exp \left[ -2 \alpha^{-1} \left( \left( \frac{r}{r_e} \right)^\alpha -1 \right) \right],
\end{equation}
where $\alpha$ is the shaping parameter, $r _e$ is the virial radius characterized by the logarithmic slope $\dv{\ln{\rho}}{\ln r} = -2$, and $\rho_e$ is the halo density measured at the virial radius $r_e$. This profile results in a Kerr-like metric in Boyer-Lindquist Coordinates of the following form
\begin{align}\label{eq:DMHaloMetric}
\begin{split}
ds^2  = -\frac{ \Delta (r) - a^2 \sin^2{\theta}}{\Sigma(r, \theta)} dt^2 + \frac{\Sigma(r, \theta)}{\Delta (r)}dr^2  +\Sigma(r, \theta) d\theta^2 \\+ 
\sin^2{\theta} \left[ a^2 + r^2 +  \frac{a \sin^2{\theta} (a^2 + r^2 - \Delta (r)) }{ \Sigma(r, \theta)} \right] d\phi^2 - \\ \frac{2 a \sin^2{\theta} (a^2 + r^2 - \Delta (r)) }{ \Sigma(r, \theta)} dt d\phi,
\end{split}
\end{align}
where the metric functions $\Delta$ and $\Sigma$ are given as follows \cite{Liu:2023oab}
\begin{subequations}
\begin{align}
\Sigma(r,\theta) &= r^2 + a^2 \cos^2\theta \label{eq:MetricFunct_Sig},\\
\Delta(r) &= r^2 g(r) - 2 M r + a^2 \label{eq:MetricFunct_Delta},\\
\begin{split}
g(r) &= \text{exp}\left\{- 4 \pi 2^{1 - 3/\alpha} e^{2/\alpha} \alpha^{3/\alpha - 1} r_e^3 \rho_e \frac{1}{r} \cdot\right.\\
&~~~~\left.\Gamma \left( 3/\alpha, 0, \frac{2}{\alpha} \left( \frac{r}{r_e} \right)^\alpha \right)\right\},\label{eq:MetricFunct_gr}
\end{split}
\end{align}
\end{subequations}
and $\Gamma$ is the incomplete generalized Gamma function. 

\begin{figure}
\centering
\includegraphics[width=1.0 \linewidth]{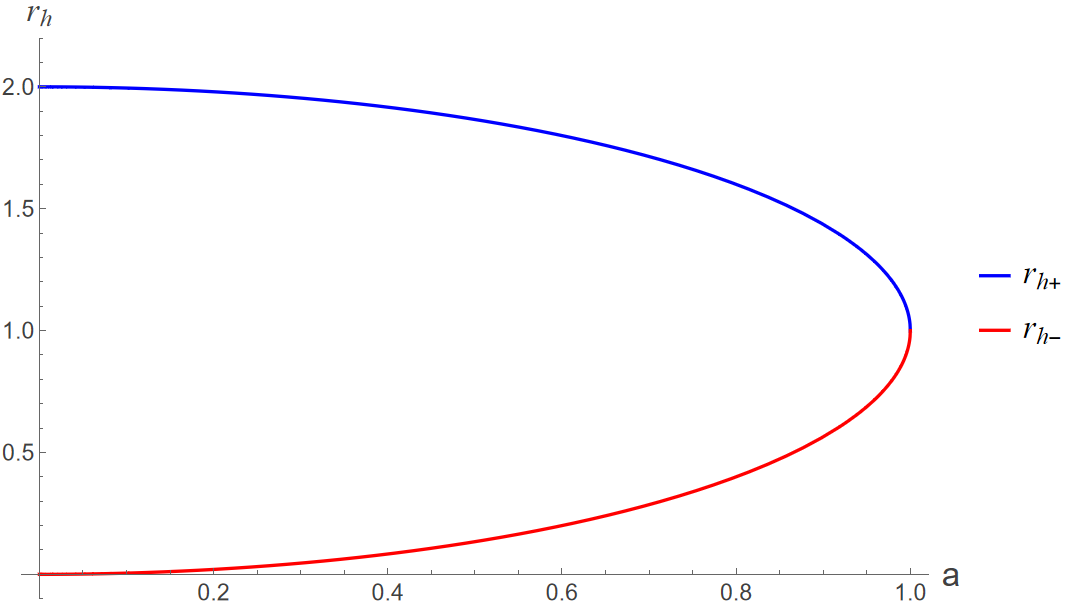}
\caption{Variation of the inner $(r_{h-})$ and outer ($r_{h+}$) horizons as a function of the black hole spin $a$. Note the derivative of both curves going to infinity at the intersection.}
\label{fig:ExtremalJustification}
\end{figure}

The multiplicative exponential term associated with $r^2$ introduces particularly interesting behavior in how the horizon structure depends on the halo parameters. Notably, setting the density parameter $\rho_e = 0$ naturally recovers the Kerr solution, without requiring adjustments to $r_e$ or $\alpha$. In the vicinity of the black hole, as $r/r_e \to 0$ (or $r\ll r_e$), we find that $g(r)\approx e^{-\gamma r^2}$, where $\gamma =\frac{8\pi}{3}e^{2/\alpha}$. This result reflects the effectively Gaussian character of the dark matter distribution in the vicinity of the black hole, where its influence is minimal. On larger, intergalactic scales, however, the parameters $\alpha$ and $r_e$ play a more significant role in shaping the overall halo profile as $r$ increases. Within the immediate gravitational environment of the black hole, variations in $\alpha$ and $r_e$ do not qualitatively alter the spacetime geometry, but instead modulate the degree to which changes in $\rho_e$ affect the horizon structure. In this paper, unless otherwise stated, we adopt the dark matter halo parameters for the M87 galaxy as $\rho_e = 6.9 E 6  M_\odot / kpc^3$, $r_e = 91.2 kpc$, and $\alpha =0.16$ \cite{Liu:2023oab}, and convert $\rho_e$ and $r_e$ into black hole units ($M=1$) for calculations in Eq.~\ref{eq:MetricFunct_gr} using
\begin{subequations}
\begin{align}
	r_e (BH units) &=  r_e / (2 G M_{BH} / c^2), \\
	\rho_e (BH units) &=  \rho_e /( M_{BH} /(4/3 \pi (2 G M_{BH} / c^2)^3 )).
\end{align}
\end{subequations}

%
%

\begin{figure}[h!]
  \begin{center}
    \includegraphics[width=\linewidth]{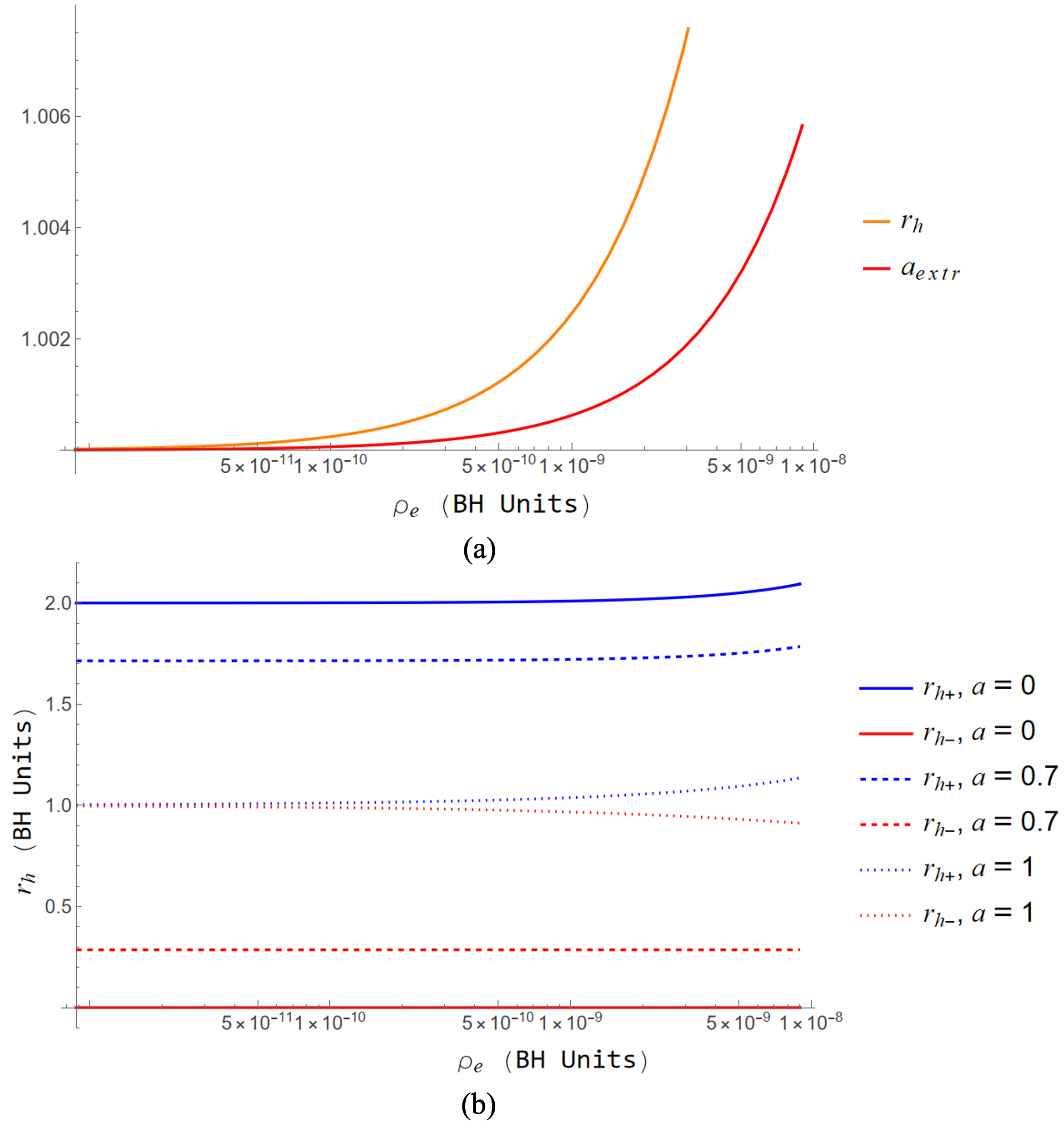}
    \caption{\label{fig:HorizonStructure} \centering Panel (a) shows $r_h$ and $a_{extr}$ with varying $\rho_e$; and panel (b) shows $r_{h+}$ and $r_{h-}$ with varying $\rho_e$. Note that both graphs are on a horizontal log scale. Also, $\rho_e \in (0, 9E-9]$ is scaled in BH units with $M=1$.}
 \end{center}
\end{figure}

One particularly notable feature of this black hole arises under extreme parameter conditions. Our analysis primarily focuses on the spin parameter $a$ and the dark matter density $\rho_e$. The spin exhibits an extremal value, denoted $a_{extr}$ at which the outer ($r_{h+}$) and inner ($r_{h-}$) horizons coincide. At this critical point, the derivatives of the horizon radii with respect to spin diverge, $\pdv{r_{h+}}{a} = \pdv{r_{h-}}{a} = \infty$, as illustrated in Fig.~\ref{fig:ExtremalJustification}. This behavior implies that the inverse derivative vanishes, $\left(\pdv{r_{h}}{a} \right)^{-1} = 0$, where the merged horizon is expressed as a function $r_h(a_{extr}, M, \rho_e, r_e, \alpha)$. This allows us to use an implicit deriviative method to find $r_h'$ from the condition $\Delta(r_h) = 0$, and then explicitly solve for $r_h$ and $a_{extr}$.

To investigate how variations in the dark matter density influence the black hole structure, we examine the dependence of $a_{extr}$ and $r_h$ on $\rho_e$, as shown in Fig.~\ref{fig:HorizonStructure}. As $\rho_e$ increases, the outer event horizon expands, consistent with the expectation that additional surrounding mass effectively deepens the gravitational potential. A noteworthy consequence of this behavior is the corresponding increase in $a_{extr}$, indicating a broader range of allowable spin values than those predicted by the Kerr metric. In realistic astrophysical contexts, however, the dark matter density in the M87 galaxy has been estimated to be on the order of $O(10^{-21})$ in black hole units, several orders of magnitude below the densities at which this shift in the extremal spin becomes appreciable.

\section{The Inclusion of Plasma in Geometrical Optics in Curved Space-time} \label{Sect:PlasmaInclusionDeriv}
In this section, we consider a more realistic scenario where the black hole is surrounded by a light-dispersive medium such as plasma, where rays no longer follow light-like geodesics of the spacetime metric. We assume no interaction between the black hole spacetime and plasma, meaning that the parameters of the rotating black hole immersed in a dark matter halo remain unaffected by plasma properties. We work within the framework of geometric optics, where light propagation is described in terms of rays rather than waves. This approach simplifies the interaction between plasma and light by considering only how the plasma affects the trajectories of the rays. Here, our focus is on studying the effect of a non-magnetized, pressureless, dust-like, and cold plasma medium, as it has a unique property that its effect on light propagation is entirely determined by the plasma frequency $\omega_p(x)$. Starting with the Hamiltonian-Jacobi equations in vacuum
\begin{equation} \label{eq:VacuumHamiltonian}
\mathcal{H} = \frac{1}{2} \tensor{p}{_\mu} \tensor{p}{^\mu} = \epsilon,
\end{equation}
where $\tensor{p}{_\mu}$ is the 4-momentum defined as $\tensor{p}{_\mu} = \tensor{g}{_\mu _\nu} \tensor{\dot{x}}{^\nu}$ and $\epsilon \in \left( -1, 0, 1\right)$ for time-like, light-like, and space-like particles, respectively. We then add the plasma contribution using the plasma frequency $\omega_p(x)^2 = (e^2/m) \mathring{n}(x)$, where $e$ is the electron charge, $m$ is the mass of the electron, and $\mathring{n}(x)$ is the electron density which is a function of our coordinates $x$ \cite{Liu:2023oab,perlick_light_2017,das_study_2022,perlick_influence_2015,chowdhuri_shadow_2021,rogers_frequency-dependent_2015}. The Hamiltonian-Jacobi equations for light-like particles is now rewritten as 
\begin{equation}\label{eq:PlasmaHamiltonian}
\mathcal{H} = \frac{1}{2} \left( \tensor{p}{_\mu} \tensor{p}{^\mu} + \omega_p(x)^2\right) = 0.
\end{equation}
\subsection{Introduction of the Index of Refraction} \label{subsect:IndexofRefract}
We proceed with the approach developed by Perlick and Tsupko in \cite{perlick_light_2017,rogers_frequency-dependent_2015} and rewrite the photon's 4-momenta $\tensor{p}{^\mu}$ relative to an observer in terms of its time-like four velocity $\tensor{U}{^\mu}(x)$:
\begin{equation}
\tensor{p}{^\mu} = - \omega(x) \tensor{U}{^\mu}(x) + \tensor{k}{^\mu}(x),
\end{equation}
where $\omega(x) = \tensor{p}{_\mu}\tensor{U}{^\mu}(x)$ and  $\tensor{k}{^\mu}(x) = \tensor{p}{^\mu} + \tensor{p}{_\nu}\tensor{U}{^\nu}(x) \tensor{U}{^\mu}(x)$ are the frequency and spatial wave vector of the light, respectively. The condition $\mathcal{H} = 0$ now becomes a dispersion relation
\begin{equation}\label{eq:DispRelat}
\omega(x)^2 = \tensor{k}{_\mu}(x) \tensor{k}{^\mu}(x) + \omega_p(x)^2,
\end{equation}
which, due to $\tensor{k}{^\mu}(x)$ being spacelike, implies that $\omega(x)^2 \ge \omega_p(x)^2$. The index of refraction can be found by simplifying Eq. \ref{eq:DispRelat}
\begin{align}\label{eq:IndexofRefraction}
\begin{split}
n(x, \omega(x))^2 \equiv \frac{1}{v_p(x, \omega(x))^2} = \frac{\tensor{k}{_\mu}(x) \tensor{k}{^\mu}(x)}{\omega(x)^2}= 1 - \frac{\omega_p(x)^2}{\omega(x)^2},
\end{split}
\end{align}
where $v_p(x, \omega(x))^2$ is the phase velocity. 

With the introduction of the refractive index, we gain a clearer physical interpretation of how plasma influences light propagation. When the photon frequency satisfies $\omega(x) > \omega_p(x)$, the refractive index is real and nonzero, allowing light to propagate through the plasma. In contrast, when $\omega(x) < \omega_p(x)$, the refractive index becomes purely imaginary, corresponding to an absorptive medium through which light cannot propagate. This condition imposes an upper limit on the plasma frequency for which transmission remains possible at a given photon frequency. Beyond this limit, the plasma surrounding the black hole becomes opaque to the photon. For real values of the refractive index, however, the plasma acts as a dispersive medium, introducing an effective lensing of photon trajectories and thereby influencing the underlying geodesic structure.

\subsection{Two Plasma Profiles} \label{subsect:PlasmaForms}
For Kerr spacetimes, it is well established in the literature \cite{perlick_influence_2015,rogers_frequency-dependent_2015,perlick_light_2017,chowdhuri_shadow_2021,das_study_2022, Liu:2023oab} that the plasma frequency has to take the following form:
\begin{equation}\label{eq:SeperablePlasFreqForm}
\omega_p(r,\theta)^2 = \frac{f_r(r) + f_\theta(\theta)}{\Sigma(r,\theta)}
\end{equation}
to allow analytically separable geodesic solutions (also explained later in Sec.~\ref{subsect:SpherGeoBH}). In this paper, we focus on studying two particular profiles of homogeneous and inhomogeneous plasma. For the homogeneous plasma case, we define
a constant electron density, i.e. $\omega_p^2=const=\omega_c^2$, where $\omega_c$ is the fundamental plasma frequency defined by $\omega_c^2 = (e^2/m) \mathring{n}$ and $\mathring{n}$ represents a uniform electron density. This yields 
\begin{subequations} 
\begin{align}
f_{r (Homo)}(r) &= \omega_c^2 r^2 \label{eq:HomoPlasmaFr}, \\
f_{\theta (Homo)}(\theta) &= \omega_c^2 a^2 \cos^2{\theta} \label{eq:HomoPlasmaFth}.
\end{align}
\end{subequations}
With this choice, $f_{r(Homo)}+f_{\theta(Homo)}=\omega_c^2\Sigma$, and therefore Eq.~\eqref{eq:SeperablePlasFreqForm} gives the intended homogeneous plasma frequency $\omega_p^2=\omega_c^2$. The advantage of adopting a homogeneous plasma profile is twofold. First, it still permits variation of the refractive index as a function of $r$ and $\theta$, since photons experience gravitational redshift while propagating through the curved spacetime \cite{perlick_light_2017}. Second, it enables an intuitive interpretation of the parameter $\omega_c$ as an effective measure of plasma density, where larger (smaller) values of $\omega_c$ correspond to stronger (weaker) plasma effects, or equivalently, greater (lesser) “plasma strength” for a given photon frequency $\omega(x)$.

For the inhomogeneous plasma profile, we adopt the combination of the two plasma distributions discussed in \cite{perlick_light_2017} and define
\begin{subequations}
\begin{align}
f_{r (Inhomo)}(r) &= \omega_c^2 \sqrt{M^3 r}, \label{eq:InhomoPlasmaFr}\\
f_{\theta (Inhomo)}(\theta) &= \omega_c^2 M^2 (1 + 2 \sin^2{\theta}) \label{eq:InhomoPlasmaFth},
\end{align}
\end{subequations}
where $M$ is the black hole mass and set to unity in black hole units. While this configuration does not represent the most complex or fully realistic plasma environment, it offers a valuable balance between physical relevance and analytical clarity. Our preliminary analysis showed that the two plasma models yield qualitatively consistent behaviors, differing mainly in the magnitude of their effects. By combining them, we enhance the visibility of plasma-induced phenomena and achieve a clearer comparison among the three cases considered: no plasma, inhomogeneous plasma, and homogeneous plasma. Moreover, this approach naturally introduces explicit dependence on both $r$ and $\theta$, aligning well with the spatial variations typically seen in GRMHD simulations \cite{gammieHARMNumericalScheme2003, theeventhorizontelescopecollaborationFirstM87Event2019b}.

\begin{figure}[h!]
  \begin{center}
    \includegraphics[width=\linewidth]{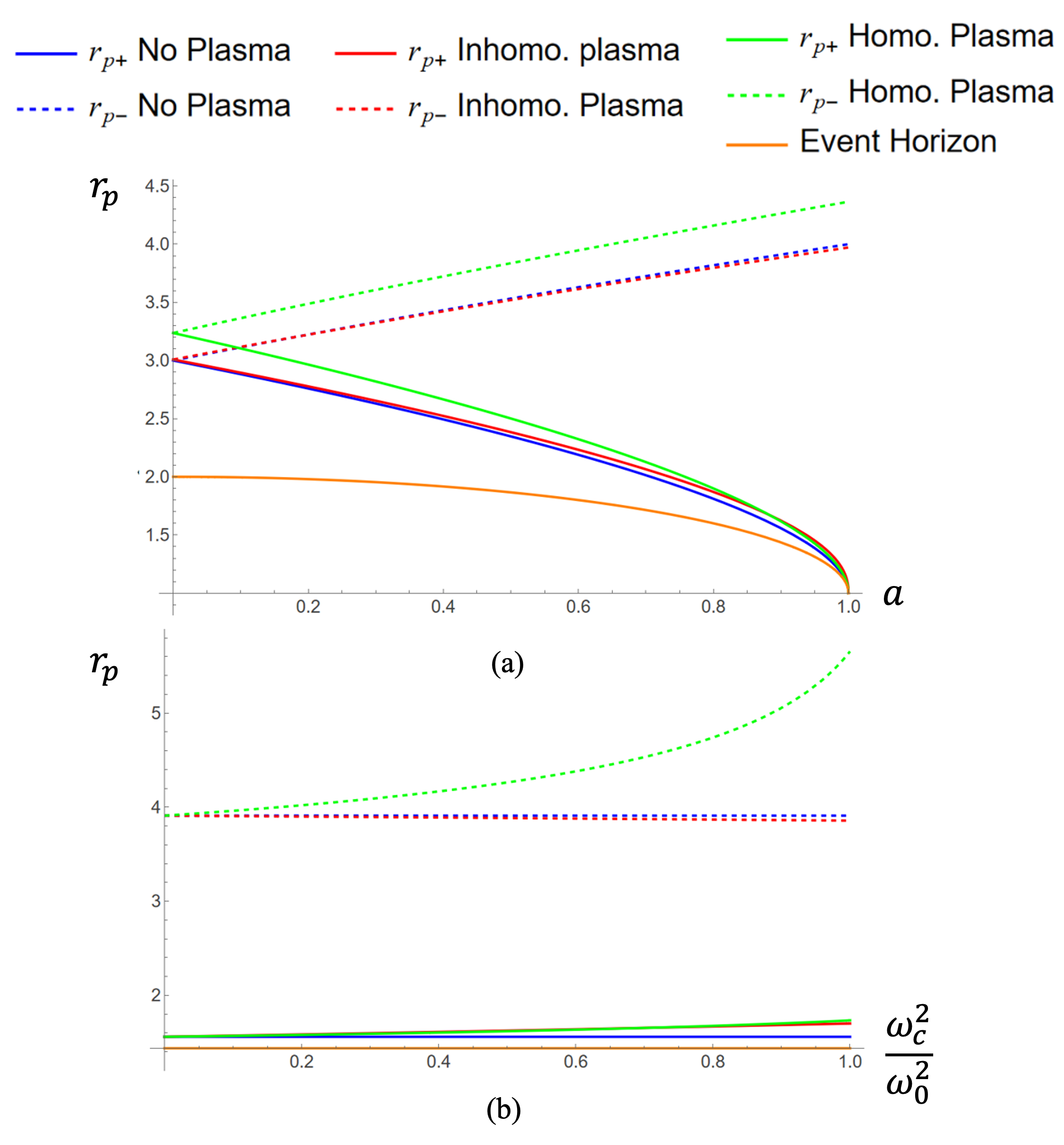}
    \caption{\label{fig:CircPhotonOrb} \centering Panel (a) shows the variation of $r_{p+}$ and $r_{p-}$ as a function of $a$ for different plasma models, while ${\omega_c^2}/{\omega_0^2}$ is kept at 0.5; Panel (b) shows the variation of $r_{p+}$ and $r_{p-}$ as a function of ${\omega_c^2}/{\omega_0^2}$ for different plasma models, while $a$ is kept at 0.99.}
 \end{center}
\end{figure}

\section{Null Geodesics and Gravitational Lensing}
\label{sect:GeoShadowDeriv}
\subsection{Plasma Influence on Photon Orbits}\label{subsect:SpherGeoBH}

\begin{figure*}[htbp]
  \begin{center}
    \includegraphics[width=7in]{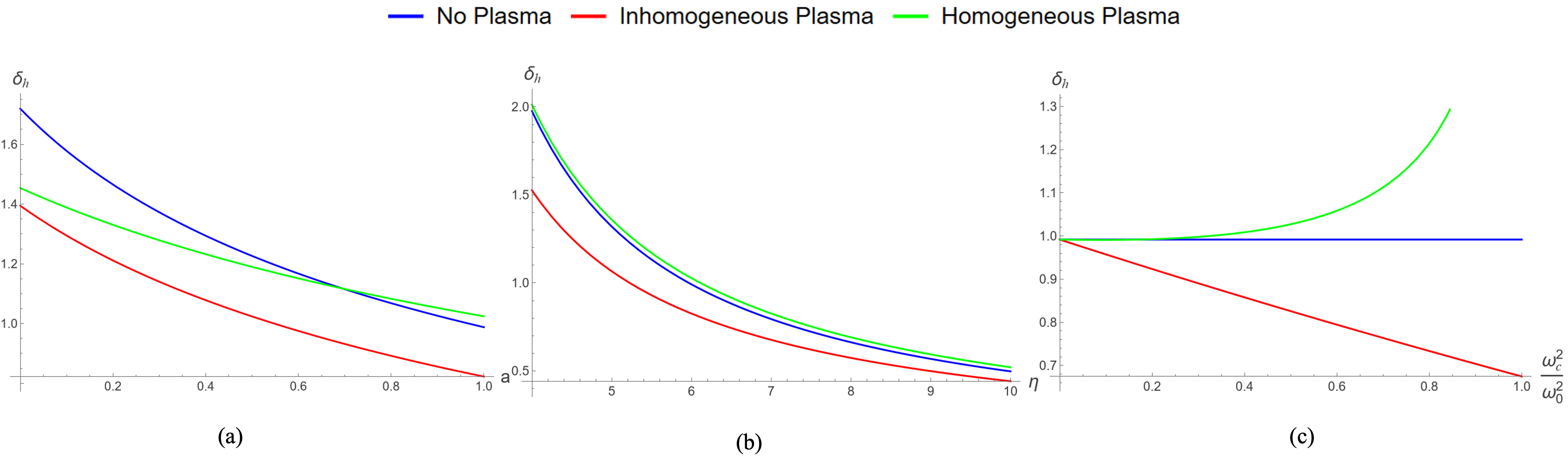}
 \caption{\label{fig:DeflectAngle} Variations of deflection angle as a function of $a$ (Panel (a)), $\eta$(Panel (b)), and ${\omega_c^2}/{\omega_0^2}$ (Panel (c)) for different plasma models. In all cases, $\rho_e = 6.9 E 6  M_\odot / kpc^3$, $r_e = 91.2 kpc$,  and $\alpha =.16$}
 \end{center}
\end{figure*}

To further study the impact of plasma on null geodesics, we assume the following separable action anzatz \cite{carter_global_1968, perlick_light_2017,das_study_2022,Liu:2023oab}
\begin{equation}
S = - E t + L_z \phi + S_r(r) + S_\theta(\theta),
\end{equation}
where $E$ and $L_z$ are the conserved energy and azimuthal angular momentum per unit mass at infinity, respectively, and $S_r(r)$, $S_\theta(\theta)$ are separated actions in terms of $r$, $\theta$, respectively. Plugging this ansatz into Eq.~\ref{eq:PlasmaHamiltonian} with $p_\mu=\frac{\partial S}{\partial x^{\mu}}$, we get

\begin{align}\label{eq:SeperableAction}
\begin{split}
\tensor{g}{^\mu ^\nu} \pdv{S}{\tensor{x}{^\mu}} \pdv{S}{\tensor{x}{^\nu}} +  \omega_p(x)^2=  \Delta(r) \pdv{S_r(r)}{r} ^2 + \pdv{S_\theta(\theta)}{\theta}^2 \\+ (L_z \csc{\theta} - a E \sin{\theta})^2 - \frac{\left((r^2 + a^2)E - a L_z\right)^2}{\Delta(r)} \\+ \Sigma(r,\theta) \omega_p(x)^2 = 0.
\end{split}
\end{align}

Clearly, for the above equation to be analytically separable $\omega_p(x)$ must take the form
 \begin{equation} \label{eq:SeperablePlasFreqAnsatz}
\omega_p(r,\theta)^2 = \frac{f_r(r) + f_\theta(\theta)}{\Sigma(r,\theta)},
\end{equation}
which leads to the following separated equations
\begin{align}
\begin{split}
\pdv{S_\theta(\theta)}{\theta}^2 + (L_z \csc{\theta} - a E \sin{\theta})^2 + f_\theta(\theta)\\ = -\Delta(r) \pdv{S_r(r)}{r} ^2 + \frac{\left((r^2 + a^2)E - a L_z\right)^2}{\Delta(r)} - f_r(r).
\end{split}
\end{align}
By introducing the Carter constant $\kappa$ and solving for the momenta, we obtain
 \begin{subequations}
 \begin{align}
\pdv{S_\theta(\theta)}{\theta} &= \sqrt{\Theta(\theta)} \\&= \sqrt{\kappa -  (L_z \csc{\theta} - a E \sin{\theta})^2 - f_\theta(\theta)}, \label{eq:ThetaActionEq}\\
\Delta(r) \pdv{S_r(r)}{r} &= \sqrt{R(r)} \\&= \sqrt{\left((r^2 + a^2)E - a L_z\right)^2 - \Delta(r) \left( \kappa + f_r(r) \right)}.\label{eq:RadialActionEq}
\end{align}
\end{subequations}
Now, using definitions $\pdv{S_r(r)}{r} = \tensor{p}{_r} = \tensor{g}{_r _r} \dot{r} = \frac{\Sigma(r,\theta)}{\Delta(r)} \dot{r}$ and $\pdv{S_\theta(\theta)}{\theta} = \tensor{p}{_\theta} = \tensor{g}{_\theta _\theta} \dot{\theta} =\Sigma(r,\theta) \dot{\theta}$, the final sphereical geodesics equations are given by
 \begin{subequations}
 \begin{align}
\left( \frac{\Sigma(r,\theta)}{E} \right) ^2 \dot{\theta} ^2 &= \frac{\Theta(\theta)}{E^2} = \chi -  (\eta \csc{\theta} - a  \sin{\theta})^2 - \tilde{f}_\theta(\theta), \label{eq:ThetaEq}\\
\left( \frac{\Sigma(r,\theta)}{E} \right) ^2 \dot{r} ^2 &= \frac{R(r)}{E^2} = \left(r^2 + a^2 - a \eta \right)^2 - \Delta(r) \left( \chi + \tilde{f}_r(r) \right), \label{eq:RadialEq}
\end{align}
\end{subequations}
where $\tilde{f}_r(r) = f_r(r)/E^2$ and $\tilde{f}_\theta(\theta) = f_\theta(\theta)/E^2$. The impact parameters $\eta = L_z/E$ and $\chi = \kappa/E^2$ introduced here can be thought of as $\phi$ and $\theta$ angular momentum conservation constants. Since the energy of a photon measured at infinity is $E = \hslash \omega_0=\omega_0$, where $\omega_0$ is the frequency of the photon at infinity, $\tilde{f}_r(r)$ and  $\tilde{f}_\theta(\theta)$ both include a $\omega_c^2/\omega_0^2$ factor, which we refer to as the analogous plasma density. To find spherical photon orbits relevant for the shadow boundary, we require $R(r_p) = 0$ and $\dv{R(r_p)}{r} =0$, i.e.
\begin{equation}
4 r_p (a^2 + r_p^2 - a \eta) - \Delta' (r_p) \left( \chi + \tilde{f}_r(r_p) \right)  - \Delta(r_p) \tilde{f}'_r(r_p) = 0.
\end{equation}
Solving this system of equations for $\eta$, $\chi$ in terms of $r_p$, we get
\begin{equation}
\begin{aligned}
\eta_\mp =& \frac{1}{a^2 \Delta'(r_p)} \left(a \left(a^2+r_p^2\right) \Delta'(r_p)-2 a r_p \Delta(r_p) \right) \\
 &+\frac{1}{a^2 \Delta'(r_p)}  \left (\mp \sqrt{a^2 \Delta(r_p)^2 \left(4 r_p^2-\tilde{f}_r'(r_p) \Delta'(r_p)\right)}\right),  \\
\chi_\pm =& \frac{1}{a \Delta'(r_p)^2}  \left ( a \Delta(r_p) \left(8 r_p^2-\tilde{f}_r'(r_p) \Delta'(r_p)\right) \right)\\
 &+\frac{1}{a \Delta'(r_p)^2}  \left(\pm 4 r_p \sqrt{a^2 \Delta(r_p)^2 \left(4r_p^2 - \tilde{f}_r'(r_p) \Delta'(r_p)\right)}\right) - \\
 &\tilde{f}_r(r_p) \label{eq:ImpactParam}.
\end{aligned}
\end{equation}
The two sign choices correspond to the two algebraic branches of the spherical-orbit conditions. To identify the branch relevant for the shadow boundary, we first take the vacuum limit $\tilde f_r,\tilde f_r'\rightarrow0$, for which Eq.~\eqref{eq:ImpactParam} reduces to
\begin{equation}
\begin{aligned}
\eta_\mp =& \frac{1}{a^2 \Delta'(r_p)} \left( a (a^2 + r_p^2) \Delta'(r_p) - 2 a r_p \Delta(r_p) \right) \\
 &+\frac{1}{a^2 \Delta'(r_p)}  \left (\mp 2 \sqrt{a^2 \Delta(r_p)^2 r_p^2 } \right),  \\
\chi_\pm =& \frac{1}{a \Delta'(r_p)^2}  \left ( 8 a r_p^2 \Delta(r_p) \right)\\
 &+\frac{1}{a \Delta'(r_p)^2}  \left(\pm  8 a r_p^2 \Delta(r_p) \right).
\end{aligned}
\end{equation}

The physical branch $\{\eta_-,\chi_+\}$ is selected by requiring continuity with the Kerr shadow boundary in this vacuum limit and real polar motion, $\Theta(\theta)\ge0$. The radial range of the unstable photon region relevant for the shadow is bounded by the equatorial plane co- and counter-rotating circular orbits \cite{perlick_calculating_2022,perlick_light_2017,bardeen_timelike_1973,Liu:2023oab,das_study_2022}. By setting $\theta=\pi/2$, we obtain $R(r)$ at the equatorial plane
\begin{gather}
\left( \frac{\Sigma(r,\theta)}{E} \right) ^2 \dot{r} ^2 = \frac{R(r)}{E^2} \\= \left(r^2 + a^2 - a \eta \right)^2 - \Delta(r) \left( (\eta - a )^2 + \tilde{f}_\theta(\frac{\pi}{2})  + \tilde{f}_r(r) \right) \label{eq:EquiRadialEq}.
\end{gather}

We numerically solve the above equation by substituting the $\{\eta_-,\chi_+\}$ from Eq.~\ref{eq:ImpactParam} and numerically solving for the co-rotating and counter-rotating photon orbits $r_p$. Fig.~\ref{fig:CircPhotonOrb} illustrates how these orbital radii vary with the black hole spin $a$ and plasma density  $\omega_c^2/\omega_0^2$. In panel (a) of Fig.~\ref{fig:CircPhotonOrb}, as the spin increases, all plasma models exhibit the expected co-rotating limiting behavior consistent with the Kerr solution, where for unit mass $r_{p+} \rightarrow 1$ as $a \rightarrow 1$, ultimately merging with the event horizon. However, unlike the Kerr case, where $r_{p-} \rightarrow 4$ for the counter-rotating orbit as $a \rightarrow 1$, our results show distinct variations among the different plasma configurations. In particular, the homogeneous plasma solution generally yields larger photon orbit radii compared to both the inhomogeneous plasma and vacuum (no plasma) cases. Additionally, the inhomogeneous plasma solution produces slightly smaller counter-rotating and slightly larger co-rotating orbits. This trend can be understood from panel (b) of Fig.~\ref{fig:CircPhotonOrb}, which shows that the counter-rotating photon orbit radius increases exponentially with plasma density in the homogeneous case, whereas it decreases approximately linearly in the inhomogeneous case. For both plasma models, the co-rotating photon orbit radius increases with plasma density, moving progressively outward from the event horizon and away from the corresponding vacuum (no plasma) orbit.
\subsection{Plasma Influence on Deflection Angle} \label{sect:DeflAng}

Next, we evaluate the deflection angle of photons with given impact parameters $(\eta,\chi)$. This angle serves as a measure of the gravitational lensing produced by the black hole. Although this calculation does not constitute a complete reconstruction of the resulting lensed image, it provides valuable insight into the behavior of light rays in the vicinity of the black hole and serves as a useful diagnostic for probing spacetime curvature and plasma effects. The deflection angle $\delta_h$ is given by
\begin{equation}
	\delta_h + \pi = \int_{R_c}^{\infty} \left. \dv{\phi}{r} \right\rvert_{\theta_0 } dr,
	\label{eq:DeflAngDef}
\end{equation}
where $R_c$ is the radius of closest approach. Applying Hamilton's equations $\tensor{\dot{x}}{^\mu} = \pdv{\mathcal{H}}{x^\mu}$, we get:
\begin{align}
\begin{split}
	\delta_h + \pi &= \int_{R_c}^{\infty} \dv{\phi}{r} dr \\&= \int_{R_c}^{\infty} \left. \frac{\frac{a}{\Delta(r)} \left[ (r^2 + a^2)  - a \eta \right] + \eta \csc^2{\theta} - a}{\pm \sqrt{\left(r^2 + a^2 - a \eta \right)^2 - \Delta(r) \left( \chi + \tilde{f}_r(r) \right)}}  \right\rvert_{\theta_0 } dr.
	\label{eq:DeflAngMidStep}
\end{split}
\end{align}
Here, $\chi$ can be solved uniquely from $\Theta(\theta)=0$ for any given $\theta_0$, which results in
\begin{equation}
\chi =  (\eta \csc{\theta_0} - a  \sin{\theta_0})^2 + \tilde{f}_\theta(\theta_0) \label{eq:chideflect}.
\end{equation}
Thus, our final equation is
\begin{align}
\begin{split}
	\delta_h + \pi &= \int_{R_c}^{\infty} \dv{\phi}{r} dr \\
	&= 2 \int_{R_c}^{\infty} \left. \frac{\frac{a}{\Delta(r)} M(r) + \eta \csc^2{\theta} - a}{\sqrt{M(r)^2 - \Delta(r) \left(\chi + \tilde{f}_r(r) \right)}}  \right\rvert_{\theta_0 } dr,
	\label{eq:DeflAngFinal}
\end{split}
\end{align}
where the factor of $2$ is due to the $\pm$ in the denominator, $\chi$ is given by Eq.~\ref{eq:chideflect}, and the subsitution $M(r) = r^2 + a^2 - a \eta$ is done for brevity.
\begin{figure*}[htbp]
  \begin{center}
    \includegraphics[width=6in]{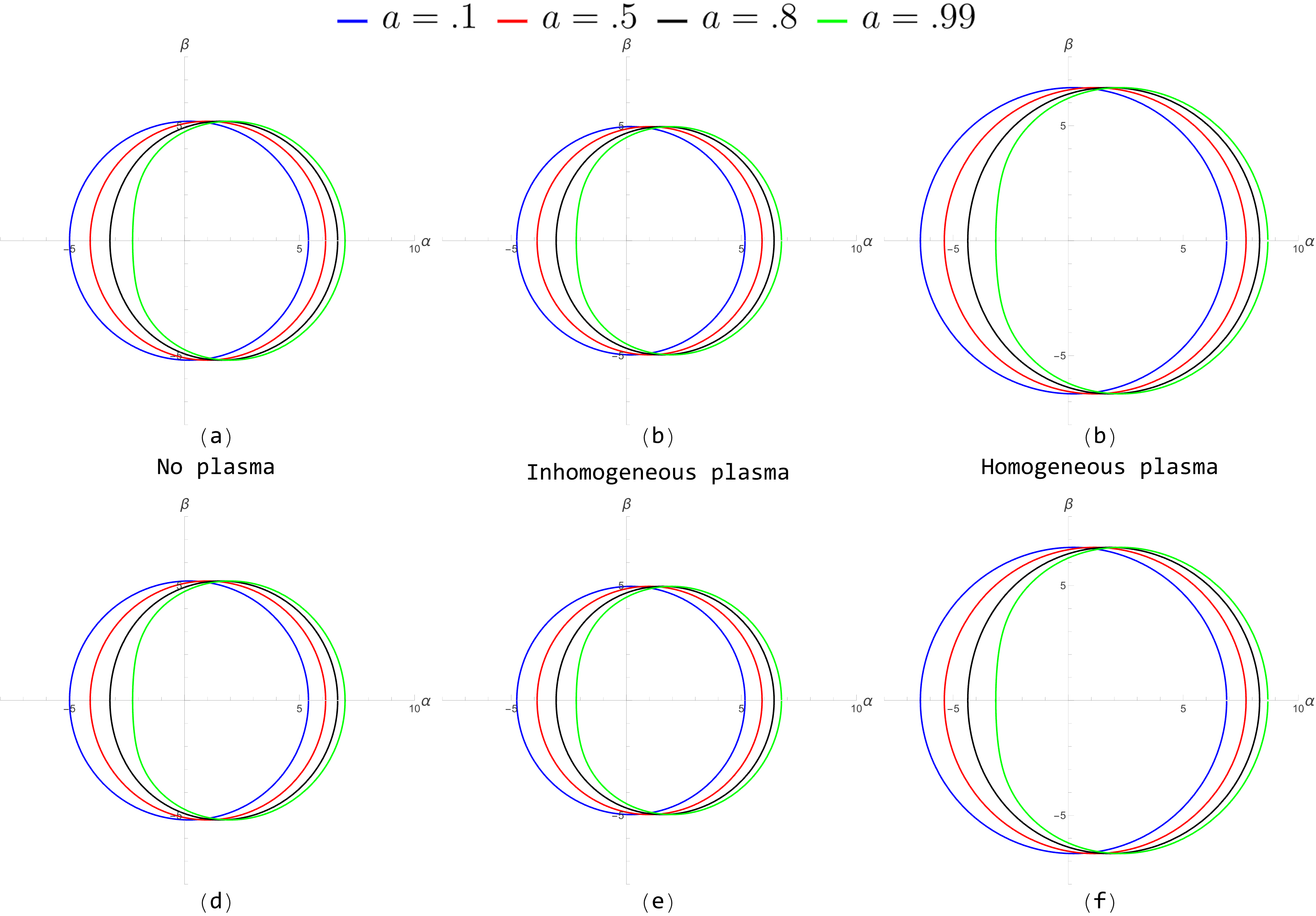}
 \caption{\label{fig:shadowshape_a} \centering Shadow shape for various values of black hole spin $a$ and inclination angles $\theta_0$ for no plasma, inhomogeneous plasma, and homogeneous plasma profiles. Panels (a)-(c): $\theta_0 = \pi/2$; panels (d)-(f): $\theta_0 = \pi/12$. In all cases, $\rho_e = 6.9 E 6  M_\odot / kpc^3$, $r_e = 91.2 kpc$, $\alpha =.16$, and $\omega_c^2/\omega_0^2 = .5$}.
 \end{center}
\end{figure*}

We numerically compute the deflection angle for varying black hole spin $a$, photon impact parameter $\eta$, and plasma density $\omega_c^2/\omega_0^2$, with the results shown in Fig.~\ref{fig:DeflectAngle}. As expected, the deflection angle decreases with increasing impact parameter $\eta$, consistent with the physical intuition that photons with larger impact parameters (or equivalently, higher angular momentum) experience weaker gravitational influence, analogous to hyperbolic trajectories in Newtonian or Keplerian dynamics. When varying the spin parameter $a$, however, distinct behaviors emerge across the different plasma models due to the presence of the $- \Delta(r) \tilde{f}_r(r)$ term in the denominator, which includes a  $- a^2 \tilde{f}_r(r)$ contribution that enhances the deflection angle. This effect is most pronounced for the homogeneous plasma model, where $\tilde{f}_r(r) \propto r^2$, resulting in a deflection angle that begins below the vacuum (no plasma) case but eventually exceeds it as $a$ increases. In contrast, the inhomogeneous plasma model, characterized by $\tilde{f}_r(r) \propto \sqrt{r}$, exhibits a more moderate variation: although its deflection angle initially lies below the no-plasma case, it decreases more slowly with increasing spin, from roughly $\approx .25$ radians to $\approx .14$ radians, than in the vacuum scenario. This behavior becomes even clearer when varying $\omega_c^2/\omega_0^2$: increasing the homogeneous plasma strength enhances the deflection angle approximately exponentially, whereas increasing the inhomogeneous plasma strength reduces it in a near-linear fashion over the range plotted.
\section{Shadow and Observables Analysis} \label{sect:ResultsAnalysis}
\subsection{Plasma Influence on Shadow} \label{subsect:ShadowCoord}

In this section, we set up celestial coordinates of the black hole shadow using Bardeen's dimensional celestial coordinates $\alpha$ and $\beta$ \cite{bardeen_timelike_1973}, as also recently used in \cite{Saha:2018zas, Liu:2023oab, das_study_2022, perlick_calculating_2022}. The celestial coordinate $\alpha$ represents the apparent perpendicular distance of the shadow boundary from the axis of black hole rotation. The coordinate $\beta$ represents the apparent perpendicular distance of the shadow boundary from its projection in the black hole's equatorial plane. For an observer at infinity, they are defined as
\begin{subequations}
\begin{align}
\alpha &= \lim_{r_0 \to \infty} - r_0^2 \sin{\theta_0} \left. \left(\dv{\phi}{r}\right)  \right\rvert_{ r_0, \theta_0 }, \\
\beta &= \lim_{r_0 \to \infty}  r_0^2 \left. \left(\dv{\theta}{r}\right)  \right\rvert_{r_0, \theta_0 },
\end{align}
\end{subequations}
where $r_0$ is the observer's radial distance from the black hole and $\theta_0$ is the observer’s viewing angle measured from the axis of black hole rotation. Applying Hamilton's equations $\tensor{\dot{x}}{^\mu} = \pdv{\mathcal{H}}{\tensor{p}{_\mu}}$, we get
\begin{subequations}
\begin{align}
\begin{split}
\alpha &= \lim_{r_0 \to \infty} - r_0^2 \sin{(\theta_0)}\cdot \\
&\left( \frac{\frac{a}{\Delta(r_0)} \left[ (r_0^2 + a^2)  - a \eta \right] + \eta \csc^2{\theta_0} - a}{\sqrt{\left(r_0^2 + a^2 - a \eta \right)^2 - \Delta(r_0) \left( \chi + \tilde{f}_r(r_0) \right)}}\right), \label{eq:GeneralAlphaEq} 
\end{split}\\
\beta &= \lim_{r_0 \to \infty}  r_0^2 \left( \frac{\sqrt{\chi -  (\eta \csc{\theta_0} - a  \sin{\theta_0})^2 - \tilde{f}_\theta(\theta_0)}}{\sqrt{\left(r_0^2 + a^2 - a \eta \right)^2 - \Delta(r_0) \left( \chi + \tilde{f}_r(r_0) \right)}}\right). \label{eq:GeneralBetaEq}
\end{align}
\end{subequations}

\begin{figure*}[htbp]
  \begin{center}
    \includegraphics[width=6in]{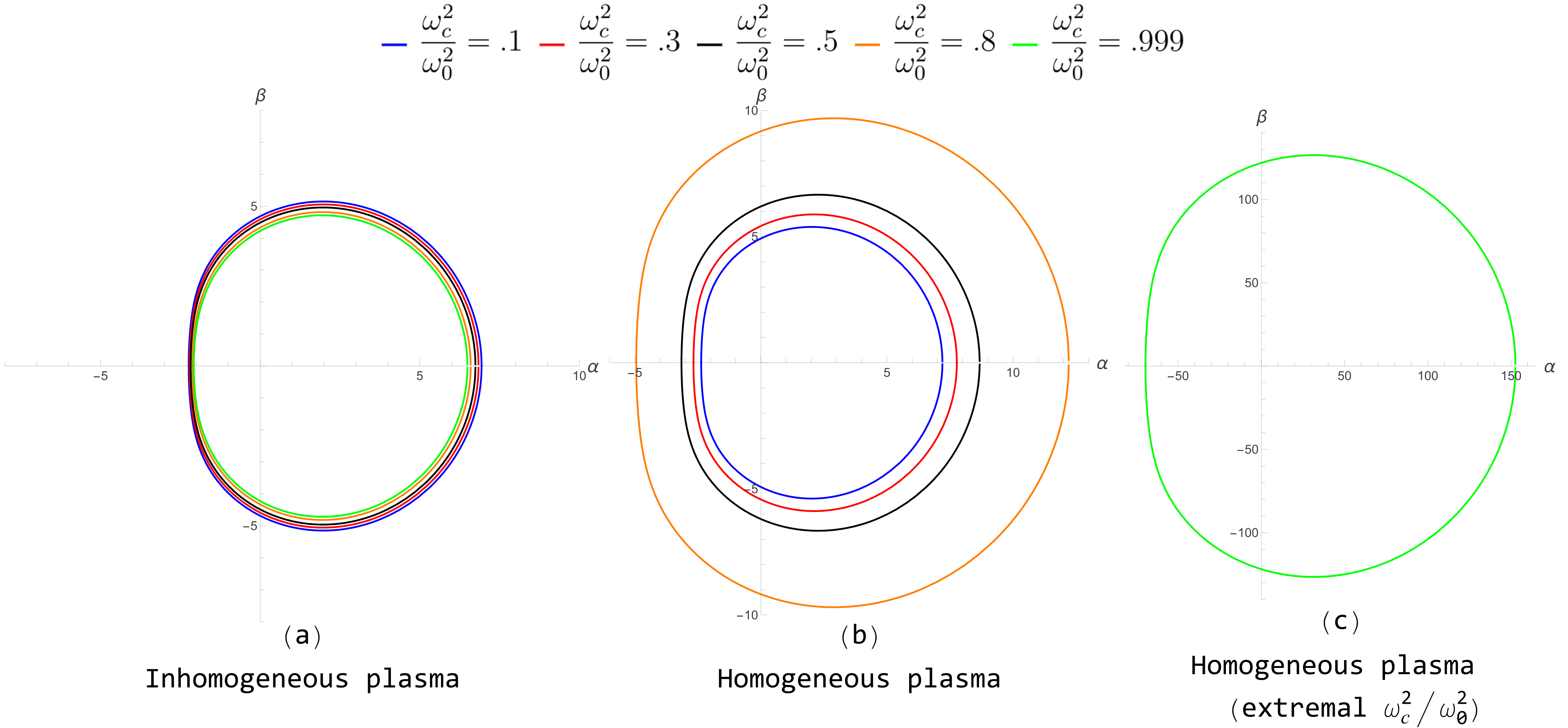}
 \caption{\label{fig:shadowshape_wc} \centering Shadow shape for various values of $\omega_c^2/\omega_0^2$ for inhomogeneous plasma and homogeneous plasma profiles. In all cases, $a=0.99$, $\rho_e = 6.9 E 6  M_\odot / kpc^3$, $r_e = 91.2 kpc$, $\alpha =.16$.}
 \end{center}
\end{figure*}

Up until this point, this analysis has been agnostic of the form of $\Delta(r)$, $\tilde{f}_r(r_0)$, and $\tilde{f}_\theta(\theta_0) $ outside of exploring the specific geodesic results of circiular photon orbits and deflection angles; however, determining their leading-order behavior in $r$ now becomes essential. From Eq.~\ref{eq:GeneralBetaEq}, we note that the numerator scales as $O(r_0^2)$, implying that for the limit $r_0 \to \infty$ to remain finite, the denominator (i.e., $\sqrt{R(r_0)}$) must also scale as $O(r_0^2)$. This condition is satisfied if $\Delta(r_0)  \tilde{f}_r(r_0) = O(r_0^4)$. Since the metric function $\Delta(r_0)$ depends on $g(r_0)$ through $r_0^2g(r_0)$, and $\lim_{r_0 \to \infty} g(r_0) = 1$ \cite{Liu:2023oab}, the leading-order behavior of $\Delta(r_0)$ is $O(r_0^2)$, constraining $\tilde{f}_r(r_0)$ to at most $O(r_0^2)$. Considering $\tilde{f}_r(r_0) = \omega_c^2 r_0^2$ for homogeneous plasma, evaluating $\sqrt{R(r_0)}$ yields a novel factor of $\sqrt{1 - (\omega_c/\omega_0)^2}$. Putting this all together, we obtain the following limits for homogeneous and inhomogeneous/no plasma, respectively,
\begin{subequations}
\begin{align}
\alpha_{Homo}  = -\frac{\eta \csc{\theta_0}}{\sqrt{1 - (\omega_c/\omega_0)^2}}, \label{eq:HomoAlphaEq} \\
\beta_{Homo}  = \frac{\sqrt{\Theta(\theta_0)}}{\sqrt{1 - (\omega_c/\omega_0)^2}}, \label{eq:HomoBetaEq}
\end{align}
\end{subequations}
\begin{subequations}
\begin{align}
\alpha_{Inhomo}  = -\eta \csc{\theta_0},\label{eq:InhomoAlphaEq} \\
\beta_{Inhomo}  = \sqrt{\Theta(\theta_0)}.\label{eq:InhomoBetaEq}
\end{align}
\end{subequations}

These relations allow us to vary the celestial coordinates with $r_p \in \left[r_{p+}, r_{p-}\right]$, corresponding to the range of the unstable spherical photon sphere bounded by the equatorial co- and counter-rotating photon orbits.
This parameterization enables us to compute and visualize the resulting black hole shadow shapes, as illustrated in Fig.~\ref{fig:shadowshape_a}. We notice several interesting trends as the viewing angle and spin parameter $a$ are varied. We observe no significant difference in the overall shadow size between the viewing angles $\theta=\frac{\pi}{2}$ and $\theta=\frac{\pi}{12}$ for all plasma models considered, as shown in Fig.~\ref{fig:shadowshape_a}. 
The shadow deformation decreases as the observer moves away from the equatorial plane, resulting in a reduced shift toward the direction of rotation (positive $a$) and less flattening on the opposite side, or a more circular shape. Additionally, for all plasma configurations, the degree of deformation increases with the black hole spin $a$; higher-spin cases exhibit stronger asymmetry, consistent with previous findings for rotating black holes \cite{perlick_light_2017,das_study_2022,Liu:2023oab}. Comparing across plasma types, we find that the inhomogeneous plasma solution generally produces slightly smaller shadows than the vacuum case, whereas the homogeneous plasma solution produces noticeably larger ones.

To further investigate plasma effects, we examine how increasing the plasma density ${\omega_c^2}/{\omega_0^2}$ influences the shadow morphology, as illustrated in Fig.~\ref{fig:shadowshape_wc}. Panel (a) shows that, in the inhomogeneous plasma model, increasing plasma density gradually reduces the overall shadow radius while leaving its deformation largely unchanged. This trend is consistent with the deflection angles shown in Fig.~\ref{fig:DeflectAngle}, where the inhomogeneous plasma produces uniformly weaker deflection compared to the other two cases, resulting in fewer photons being captured by the black hole and thus a smaller shadow. In contrast, panel (b) shows that the homogeneous plasma exhibits the opposite behavior: the shadow size increases dramatically with plasma strength. This arises from the scaling relation discussed in Sec.~\ref{subsect:ShadowCoord}, where the celestial coordinates are multiplied by a factor of $1 / \sqrt{1 - (\omega_c^2 / \omega_0^2)}$. As ${\omega_c^2}/{\omega_0^2}\rightarrow 1$, this factor diverges, producing a pronounced inflation in shadow size, as shown in Fig.~\ref{fig:shadowshape_wc}(c).

%
%
\subsection{Shadow Observables} \label{subsect:ShadowObserv}

\begin{figure}[h]
\centering
\includegraphics[width=0.9\linewidth]{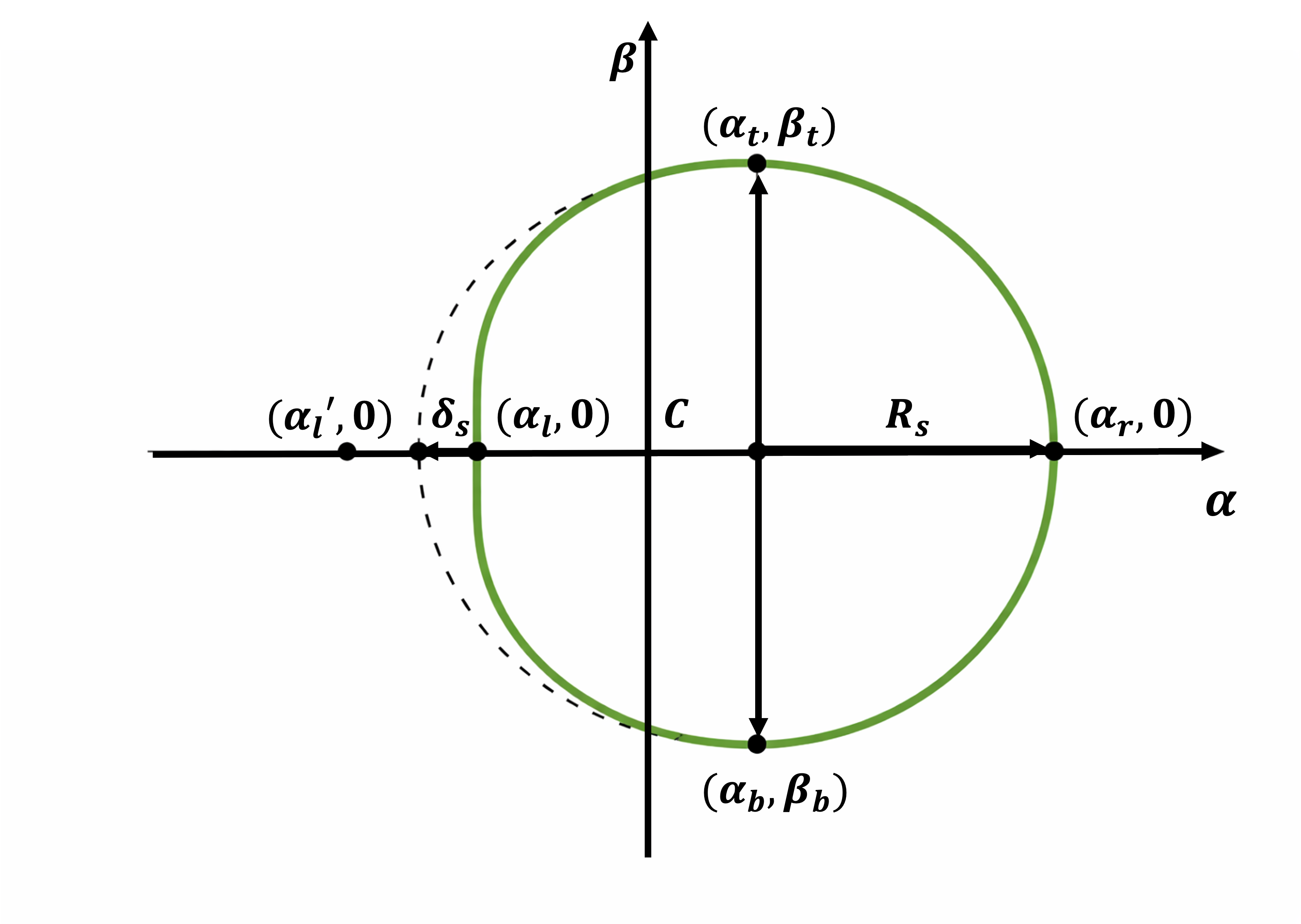}
\caption{Diagram of the shadow observables with relevant points labeled.}
\label{fig:ShadowObservDiagram}
\end{figure}

Beyond the overall shape, several measurable quantities can be used to characterize the black hole shadow, including its radius, deformation, and ellipticity \cite{Hioki:2009na, Abdujabbarov:2015xqa, Tang:2022uwi, Liu:2023oab}. The shadow can be approximated as a nearly circular contour with an effective radius $R_s$, while deviations from circularity are quantified by the flattening deformation $\delta_s$ and the ellipticity $K_s$, defined as the ratio of the vertical to horizontal widths. These quantities are derived from the characteristic points shown in Fig.~\ref{fig:ShadowObservDiagram}, where the reference circle passing through the three points $(\alpha_t, \beta_t)$, $(\alpha_r, 0)$, and $(\alpha_b, \beta_b)$, defines the effective shadow radius $R_s$ and the center position $(\alpha_c,0)$:
\begin{subequations}
\begin{align}
	R_s = \frac{(\alpha_r - \alpha_t)^2 + \beta_t^2}{2 \abs{\alpha_r - \alpha_t}}, \\
	\alpha_c = \frac{1}{2} \left( \alpha_r + \alpha_t + \frac{\beta_t^2}{-\alpha_r + \alpha_t} \right).
\end{align}
\end{subequations}
We also make use of the reflectional symmetry of the Kerr-like shadow about the $\alpha$-axis, which implies $(\alpha_b, \beta_b) =(\alpha_t, - \beta_t)$. Using this symmetry, the deformation parameter $\delta_s$ can be defined as the ratio between the deviation of the observed leftmost edge of the shadow from the reference circle and the effective shadow radius $R_s$. Specifically, it is given by the difference between the theoretical leftmost coordinate $\alpha_l'$ of the reference circle and the corresponding measured coordinate $\alpha_l$ of the Kerr-like shadow, normalized by $R_s$
\begin{equation}
	\delta_s = \frac{\alpha_l - \alpha_l'}{R_s} = \frac{\alpha_l - (\alpha_c - R_s)}{R_s}. 
\end{equation}
And the ellipticity $K_s$, as stated above, is defined as the ratio of the horizontal width $\Delta \alpha$ and $\Delta \beta$:
\begin{equation}
	K_s = \frac{\Delta \beta}{\Delta \alpha} = \frac{2 \beta_t}{\alpha_r - \alpha_l} .
\end{equation}

\begin{figure*}[htbp]
  \begin{center}
    \includegraphics[width=7in]{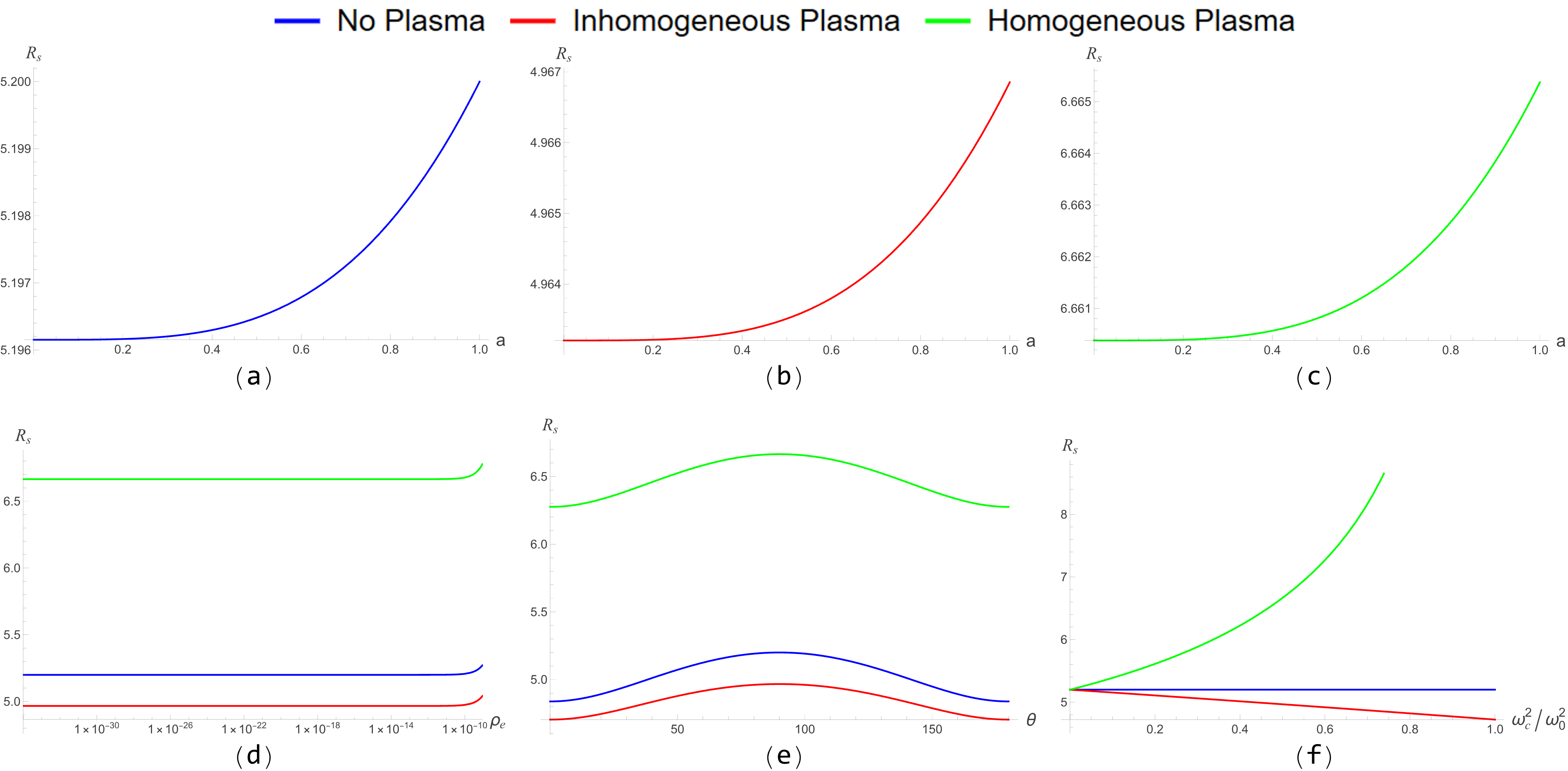}
 \caption{\label{fig:ShadowRadius} \centering Panels (a)-(c) show the variation of shadow radius \centering $R_s$ as a function of black hole spin a for different forms of plasma. We set $\theta_0 = \pi/2$, ${\omega_c^2}/{\omega_0^2}=0.5$, $\rho_e = 6.9 E 6  M_\odot / kpc^3$,  $r_e = 91.2 kpc$, and $\alpha =.16$. Panel (d) shows the variation of shadow radius as a function of the log of the dark matter halo density $\rho_e$,  while ${\omega_c^2}/{\omega_0^2}=0.5$, $\theta_0=\pi/2$, and $a=0.99$. Panel (e) shows the variation of shadow radius as a function of viewing angle $\theta$, while $\theta_0 = \pi/2$ and $a=0.99$. Panel (f) shows the variation of shadow radius as a function of ${\omega_c^2}/{\omega_0^2}$, while $\theta_0 = \pi/2$ and $a=0.99$.}
 \end{center}
\end{figure*}

We first study the dependence of the shadow radius on the spin $a$, observer inclination angle $\theta_0$, plasma density ${\omega_c^2}/{\omega_0^2}$, and dark matter density $\rho_e$, as illustrated in Fig.~\ref{fig:ShadowRadius}. We observe that the shadow size depends only weakly on spin (a slight growth of $\approx .07\%$) across all plasma models, but decreases gradually as the viewing angle $\theta_0$ approaches its limits of 0 and $\pi$, as shown in panels (a)-(c) and (e), respectively. In addition, the plasma density ${\omega_c^2}/{\omega_0^2}$ affects the shadow radius differently depending on the plasma profile: the homogeneous plasma model produces an exponentially increasing shadow radius relative to the vacuum case, whereas the inhomogeneous model shows an approximately linear decrease below the vacuum value, as illustrated in panel (f). We also explore the dependence of the shadow radius on the dark matter density $\rho_e$ as we increase it from $10^{-34}$ to $10^{-9}$ in BH units in panel (d). Notice that the radius does not change significantly until $\rho_e > 10^{-11}$, where our values for M87* and Sgr A* are on the order of $10^{-21}$ and $10^{-28}$ BH units, respectively, which implies that for most physical dark matter densities, there is no significant change in the shadow radius.

\begin{figure*}[htbp]
  \begin{center}
    \includegraphics[width=7in]{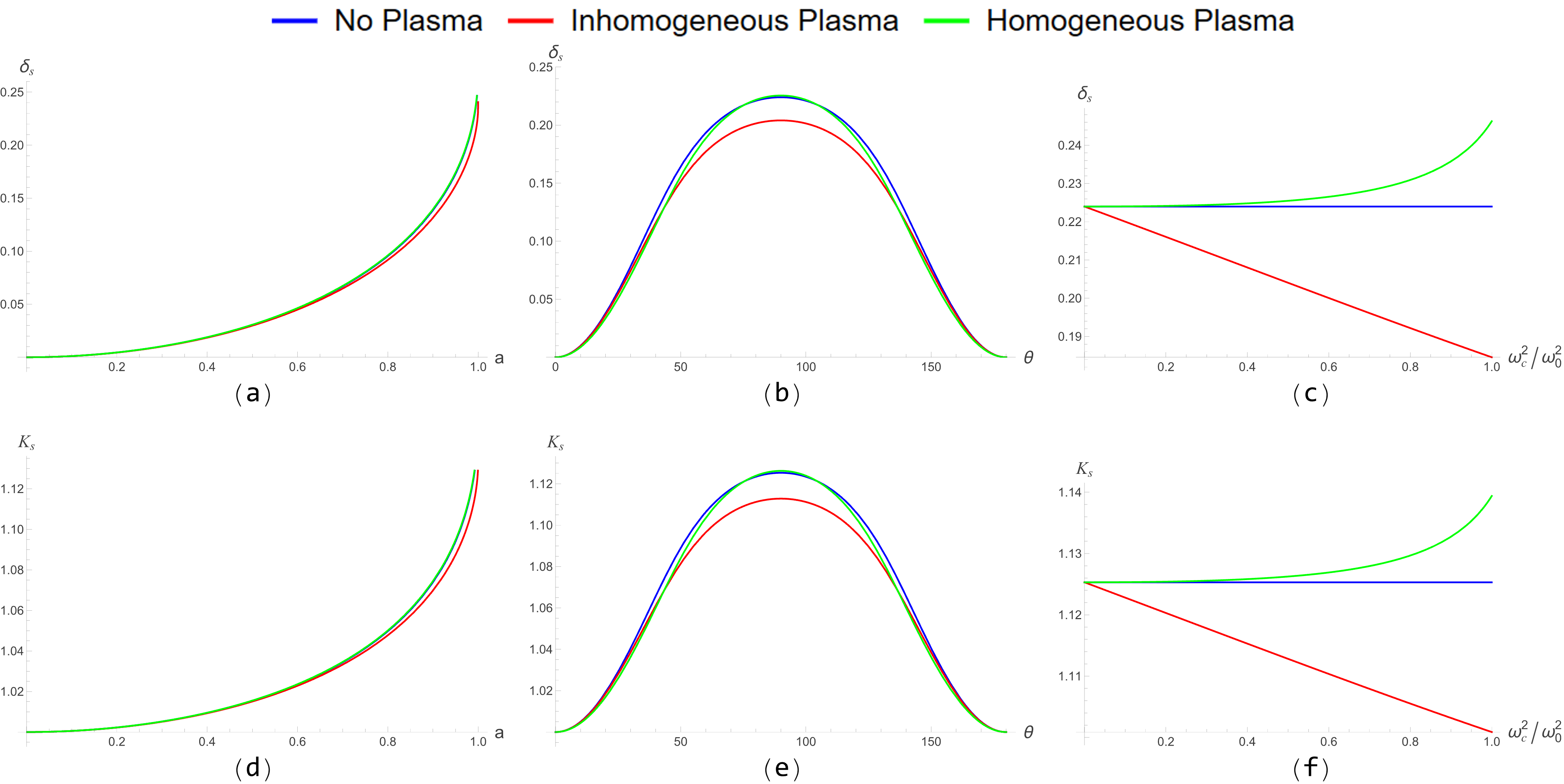}
 \caption{\label{fig:ShadowDeformEllip} \centering Panels (a)-(c) show the variation of deformation $\delta_s$ as a function of black hole spin a, viewing angle $\theta$, and plasma frequency ${\omega_c^2}/{\omega_0^2}$ for different forms of plasma. Panels (d)-(f) show the variation of ellipticity $K_s$ as a function of black hole spin a, viewing angle $\theta$, and plasma frequency ${\omega_c^2}/{\omega_0^2}$ for different forms of plasma. For Panels (a) and (d), we set $\theta_0 = \pi/2$, and $\omega_c^2/\omega_0^2 = 0.5$; for Panels (b) and (e), we set $a = 0.99$, and $\omega_c^2/\omega_0^2 = 0.5$; for Panels (c) and (f), we set $a = 0.99$, and $\theta_0 = \pi/2$.}
 \end{center}
\end{figure*}


\begin{figure}[htbp]
  \begin{center}
    \includegraphics[width=\linewidth]{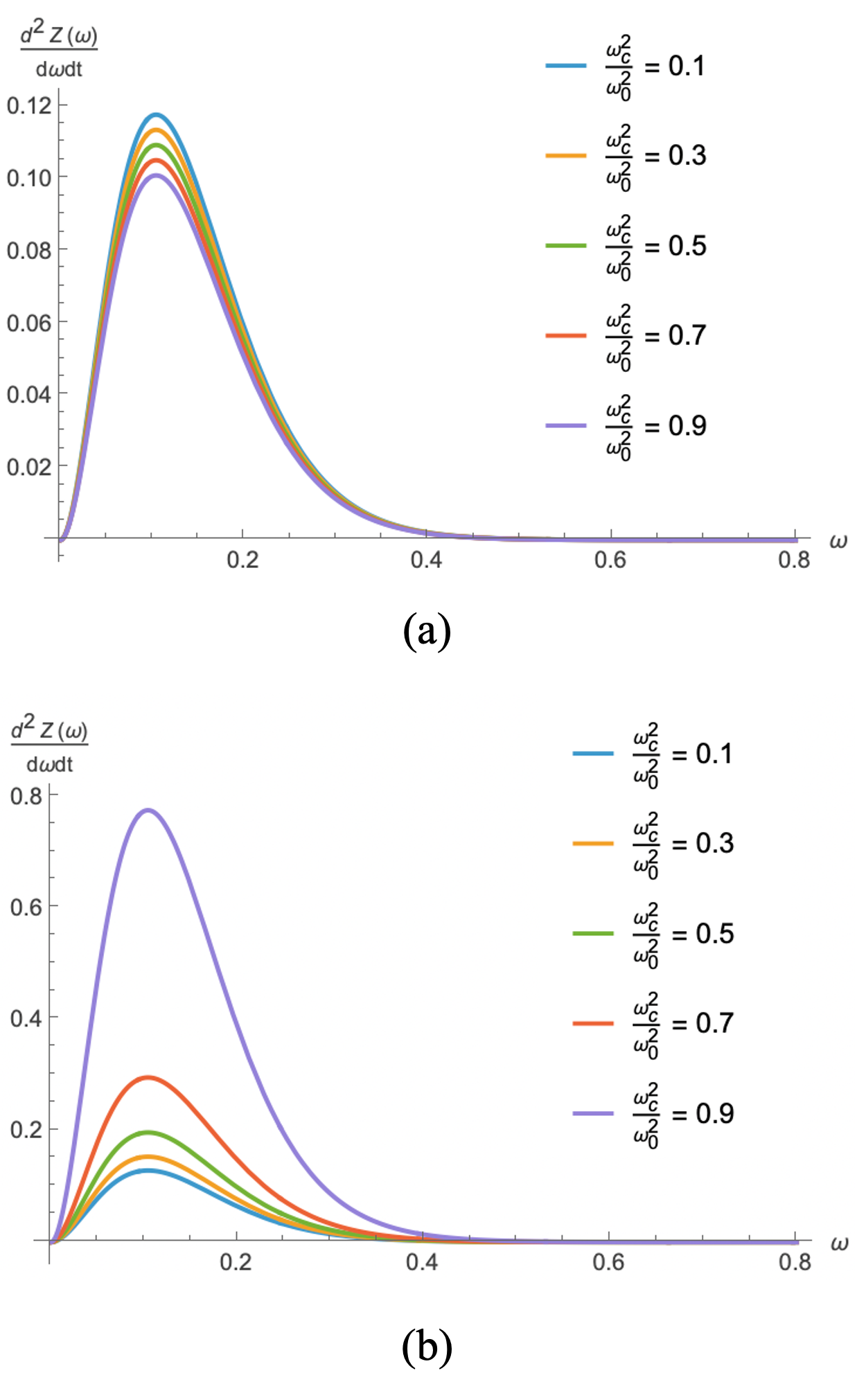}
    \caption{\label{fig:em} \centering (a) Energy emission rate $\frac{d^2Z}{d\omega dt}$ as a function of frequency $\omega$ for inhomogeneous plasma; (b) Energy emission rate $\frac{d^2Z}{d\omega dt}$ as a function of frequency $\omega$ for homogeneous plasma. In both cases, $a = 0.5$, $\rho_e = 6.9 E 6  M_\odot / kpc^3$,  $r_e = 91.2 kpc$, and $\alpha =.16$. }
 \end{center}
\end{figure}

\begin{figure*}[htbp]
  \begin{center}
    \includegraphics[width=7in]{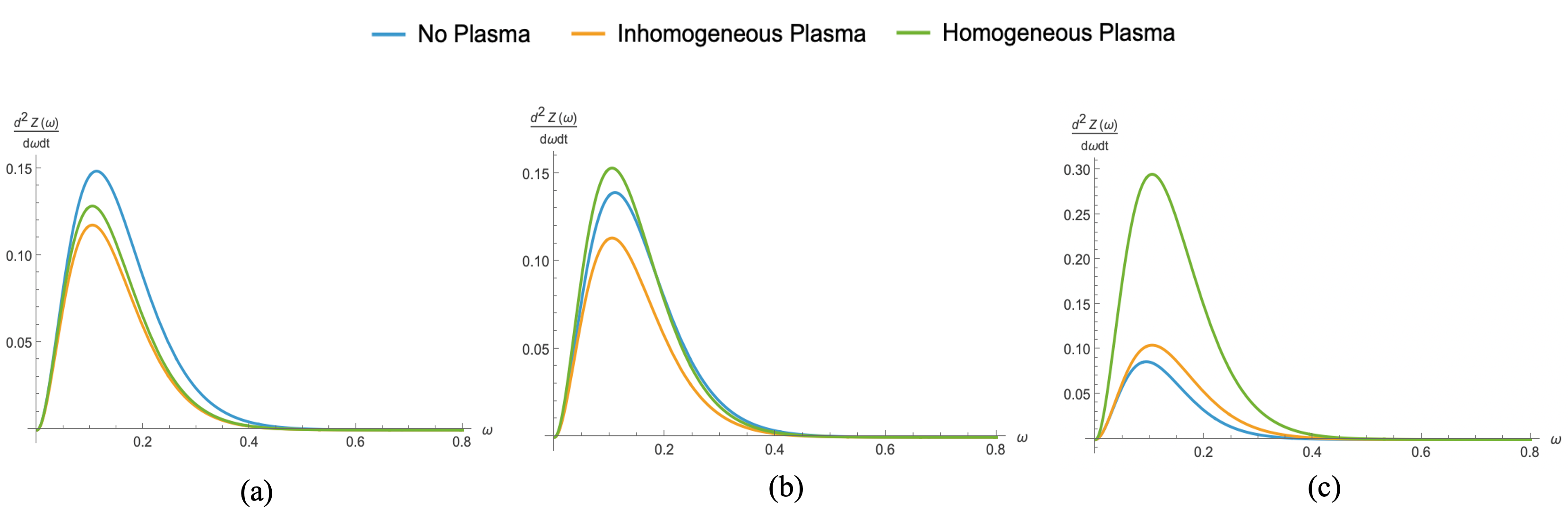}
    \caption{\label{fig:em-varfreq} \centering Energy emission rate $\frac{d^2Z}{d\omega dt}$ as a function of frequency $\omega$ for different plasma profiles: (a) ${\omega_c^2}/{\omega_0^2} = 0.1$; (b) ${\omega_c^2}/{\omega_0^2} = 0.3$; (c) ${\omega_c^2}/{\omega_0^2} = 0.7$. In all cases, $a = 0.5$, $\rho_e = 6.9 E 6  M_\odot / kpc^3$,  $r_e = 91.2 kpc$, $\alpha =.16$.  }
 \end{center}
\end{figure*}

For the deformation $\delta_s$ and ellipticity $K_s$, panels (a) and (d) in Fig.~\ref{fig:ShadowDeformEllip} show that they both grow as the black hole spin $a$ increases in nearly identical trends, consistent with the increasing flattening of the shadow observed at high spin in Fig.~\ref{fig:shadowshape_a}. Interestingly, the homogeneous plasma model remains closely aligned with the vacuum (no plasma) case, unlike the shadow radius behavior in Fig.~\ref{fig:ShadowRadius}, while the inhomogeneous plasma model shows a general decrease in both deformation and ellipticity with increasing spin. A similar pattern arises when varying the observer inclination angle $\theta_0$ shown in panels (b) and (e): the homogeneous plasma case closely tracks the vacuum case, whereas the inhomogeneous plasma case produces smaller deformation and ellipticity values. As the viewing angle approaches its limits, the effects of plasma models on $\delta_s$ and $K_s$ become indistinguishable, and the shadow becomes more circular. Finally, when varying the plasma density ${\omega_c^2}/{\omega_0^2}$, panels (c) and (f) illustrate that the deformation $\delta_s$ and ellipticity $K_s$ for the homogeneous plasma remain close to the vacuum case up to approximately ${\omega_c^2}/{\omega_0^2} = .5$, beyond which they diverge exponentially. In contrast, $\delta_s$ and $K_s$ for the inhomogeneous plasma decrease roughly linearly with increasing plasma density, consistent with its lower deformation and ellipticity values in the previous figures.

\section{Energy Emission Rate}\label{sect:EERate}

The energy emission of black holes, most notably through Hawking radiation, arises from a process in which virtual particle pairs form near the event horizon, with one particle escaping to infinity while its partner falls inward, leading to a net loss of mass for the black hole. In this section, we are interested in investigating how various plasma profiles alter the energy emission rate in the spacetime of the Kerr-like black hole in a dark matter halo under consideration. The energy emission rate can be generally expressed as \cite{das_shadow_2020}
\begin{equation}
    \frac{d^2Z(\omega)}{d\omega dt} = \frac{2 \pi^2 \sigma_{lim}}{e^{\frac{\omega}{T_H}} -1 }\omega^3,
\end{equation}
where $\sigma_{lim}$ is the limiting constant, $T_H$ is the Hawking temperature, and $\omega$ is the frequency. The limiting constant, which describes the effective area of the black hole shadow, can be expressed as
\begin{equation}
    \sigma_{lim}=\frac{\pi^\frac{d-2}{2}R_s^{d-2}}{\Gamma(\frac{d}{2})},
\end{equation}
where $R_s$ is the effective radius of the shadow and $d$ is the dimension. In $d = 4$ dimensions, we have
\begin{equation}
    \sigma_{lim} = \pi R_s^2.
\end{equation}
The Hawking temperature is given by \cite{page_hawking_2005}
\begin{equation}
    T_H = \frac{\kappa}{2\pi},
\end{equation}
where $\kappa$ is the surface gravity defined as in \cite{chou_radiating_2020} using inner horizon $r_{h-}$, outer horizon $r_{h+}$ and spin $a$:
\begin{equation}
    \kappa = \frac{r_{h+}-r_{h-}}{2(r_{h+}^2 + a^2)}.
\end{equation}

Fig.~\ref{fig:em} shows the energy emission rate $\frac{d^2Z(\omega)}{d\omega dt}$ as a function of frequency $\omega$ for different plasma frequencies in the inhomogeneous and homogeneous plasma distributions. We see that for the inhomogeneous plasma model, the emission rate decreases as the plasma frequency increases, whereas in the homogeneous plasma model, it increases. We also compare the emission rates across plasma profiles with different plasma frequencies in Fig.~\ref{fig:em-varfreq}. At low plasma frequency (${\omega_c^2}/{\omega_0^2} = 0.1$), the vacuum case produces the highest energy emission rate. The inhomogeneous plasma gives the lowest emission, with the homogeneous case falling in between. As the plasma frequency increases, the homogeneous plasma case has a much stronger effect, pushing the emission rate higher compared to both the inhomogeneous plasma and no-plasma scenarios. At high plasma frequency (${\omega_c^2}/{\omega_0^2} = 0.7$), the inhomogeneous case exhibits a higher emission rate than the vacuum case. The change in ordering mirrors how each plasma profile modifies the shadow radius $R_s$, as we have seen in Fig.~\ref{fig:ShadowRadius}, which determines the effective area used in computing the energy emission rate.

\section{Geometric Comparison with EHT-Inferred Shadow Sizes of M$87^*$ and $\text{Sgr A}^*$}\label{sect:CEHTM87As}

In 2019, the Event Horizon Telescope (EHT) collaboration released the first horizon-scale image of the supermassive black hole in the galaxy M87*, providing the first direct observational evidence of a horizon-scale compact emission region associated with a black hole \cite{EHT_first_2019}. This image revealed a bright emission ring surrounding a central intensity depression, whose angular diameter was shown to be consistent with the theoretical expectations for the shadow cast by a Kerr black hole. The observed ring size was inferred through model-dependent fits to the interferometric data and was found to be robust against variations in the underlying emission models. In 2022, the EHT collaboration reported the first image of the accretion flow surrounding Sgr~A$^*$, extending horizon-scale imaging to a second system with vastly different mass, accretion environment, and variability timescale \cite{EventHorizonTelescope:2022urf, ciurlo_sagittarius_2025}. These observations provide a useful benchmark for geometric shadow calculations, but they do not directly measure the mathematical shadow boundary. The measured quantity is a brightness ring and central depression produced by a specific emitting plasma, while our calculation concerns the critical curve for photon propagation through an external refractive plasma. Therefore, the comparison below should be read only as an illustrative geometric benchmark, not as a full EHT inference or measurement of the plasma density.

For a theoretical critical curve, the apparent size on the observer’s sky is determined by the lensing of light near unstable photon orbits. To facilitate comparisons with theoretical models and to minimize dependence on detailed shadow distortions, we characterize this size in terms of an effective (areal) shadow radius, denoted $r_{sh}$. This quantity is defined through the total area $A_{sh}$ enclosed by the shadow boundary as
\begin{equation}
r_{sh} \equiv \sqrt{\frac{A_{sh}}{\pi}} .
\end{equation}
By construction, $r_{sh}$ represents the radius of a circular disk with the same area as the theoretical critical curve and provides a robust characterization of the shadow size. Importantly, $r_{sh}$ is an apparent celestial radius measured at infinity and should not be confused with the photon-sphere radius or any coordinate radius in the spacetime.

\begin{figure}[h!]
  \begin{center}
    \includegraphics[width=\linewidth]{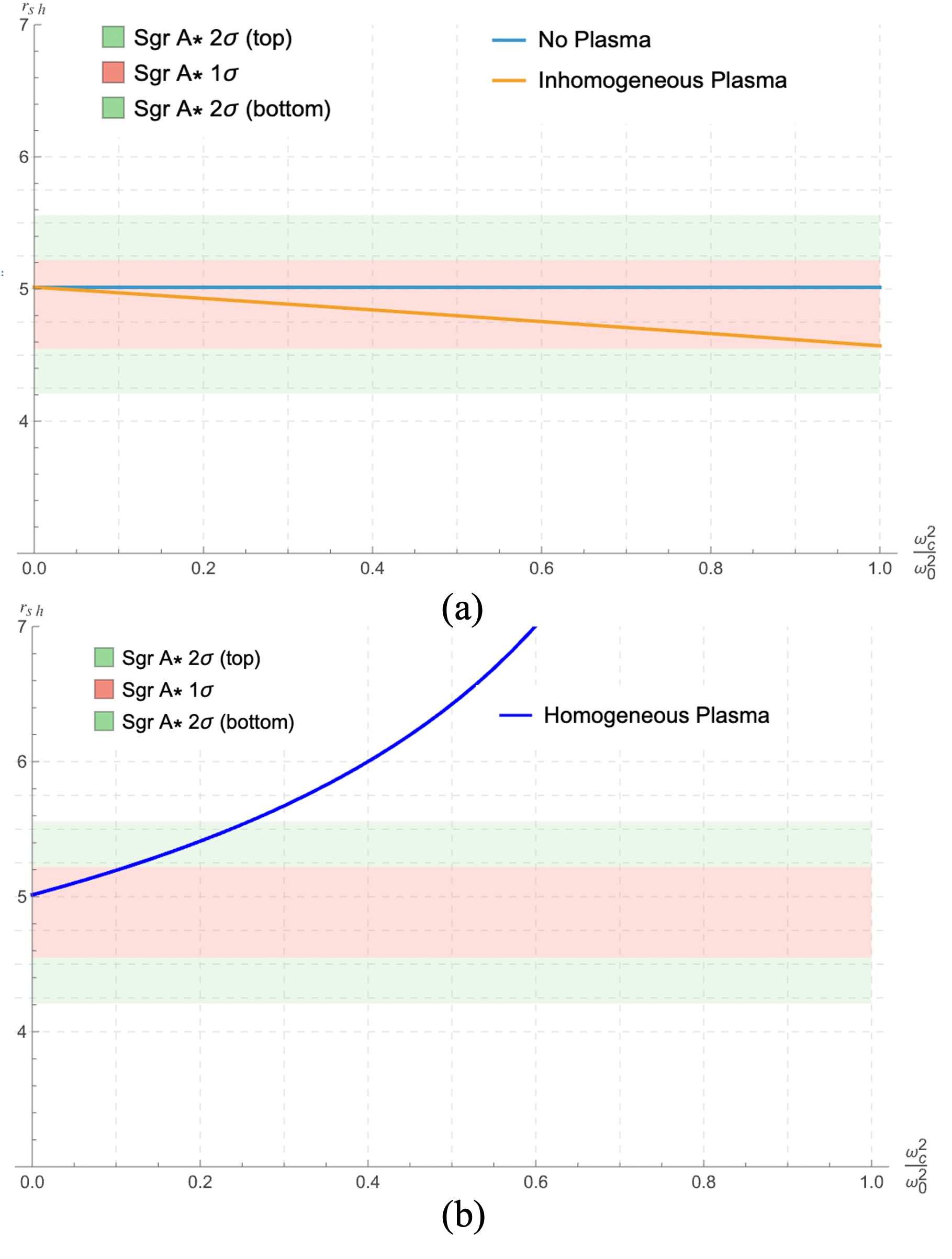}
    \caption{\label{fig:sgr} \centering Illustrative comparison of the geometric shadow radius for various plasma profiles with the EHT-inferred effective shadow-size interval of $\text{Sgr A}^*$. Here, we set $a = 0.9$, $\rho_e = 1.1 E 7  M_\odot / kpc^3$,  $r_e = 12 kpc$ and $\theta_0\approx \pi/2$.}
 \end{center}
\end{figure}

\begin{figure}[h!]
  \begin{center}
    \includegraphics[width=\linewidth]{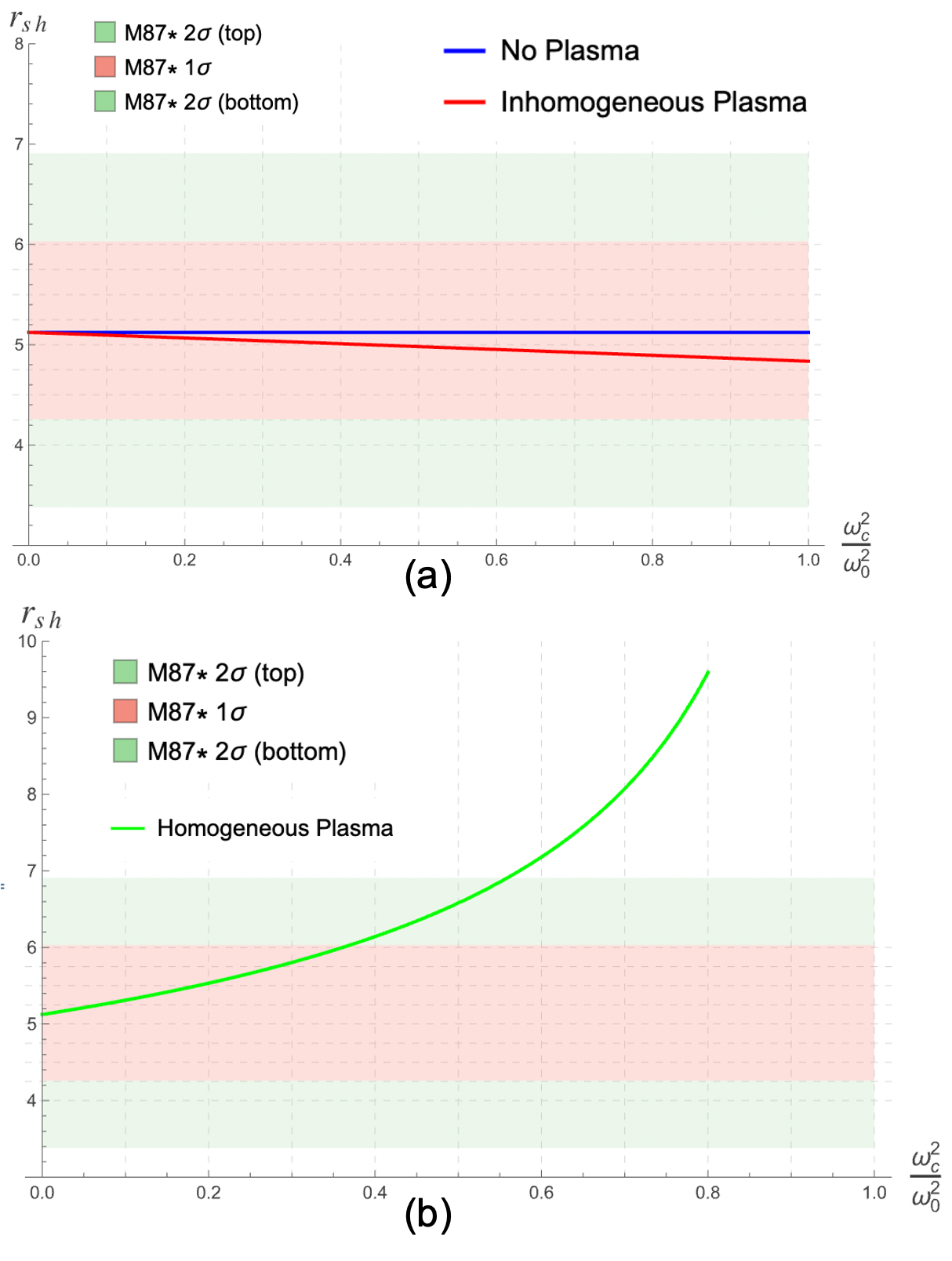}
    \caption{\label{fig:M87} \centering Illustrative comparison of the geometric shadow radius for various plasma profiles with the EHT-inferred effective shadow-size interval of M$87^*$. Here, we set $a = 0.5$, $\rho_e = 6.9 E 6  M_\odot / kpc^3$,  $r_e = 91.2 kpc$, $\alpha =.16$ and $\theta_0\approx 17^\circ.$}
 \end{center}
\end{figure}

To compare different black hole systems, the shadow size is typically rendered dimensionless by normalizing with the black hole mass, $r_{sh}/M$, where $M$ is expressed in geometric units. Observationally, this normalization is achieved by combining the measured angular size with independent estimates of the black hole mass and distance, which motivates a comparison between EHT-inferred size intervals and theoretical predictions. Because this comparison depends on the mapping between the image ring and the critical curve, we do not treat it as a direct constraint on the plasma model. In practice, the shadow boundary extracted from ray-tracing simulations or observational reconstructions is not perfectly circular, especially for rotating black holes viewed at finite inclination. As a result, the shadow can be characterized using its horizontal and vertical extents, defined respectively by

\begin{equation}
R_h=\frac{\alpha_r - \alpha_l}{2}, ~~ R_v=\frac{\beta_t-\beta_b}{2},
\end{equation}
where $(\alpha_l, 0)$ and $(\alpha_r,0)$ denote the leftmost and rightmost points of the shadow on the observer’s screen, and $(0, \beta_t)$ and $(0, \beta_b)$ denote the top and bottom points. When the shadow is approximated as an ellipse with semi-axes $R_h$ and $R_v$, its area may be written as $A_{sh}=\pi R_h R_v$, leading to the widely used approximation \cite{Abdujabbarov:2015xqa}
\begin{equation}
r_{sh} \approx \sqrt{R_h R_v}.
\end{equation}
This expression provides an efficient and accurate proxy for the areal shadow radius in Kerr-like spacetimes, with deviations typically at the percent level even for rapidly rotating black holes. In Figs.~\ref{fig:sgr} and \ref{fig:M87}, and in Tables~\ref{tab:wc-ranges-sgr} and \ref{tab:wc-ranges-M87}, we use this ellipse-proxy radius $r_{sh}\simeq\sqrt{R_hR_v}$ rather than the reference-circle radius $R_s$ introduced in Sec.~\ref{subsect:ShadowObserv}. The latter is used in Fig.~\ref{fig:ShadowRadius} and in the energy-emission estimate. The two definitions agree closely for the nearly circular cases considered here, but we keep the notation distinct to avoid ambiguity.
For both sources, we convert the fractional shadow-size deviation $\delta$ to an effective dimensionless radius through
\begin{equation}
\frac{r_{sh}}{M}=3\sqrt{3}\,(1+\delta).
\end{equation}
For M$87^*$, Ref.~\cite{EventHorizonTelescope:2021dqv} reports $\delta_{\rm M87^*}=-0.01\pm0.17$ at $68\%$ confidence. This gives the $1\sigma$ interval below, while the quoted $2\sigma$ interval is obtained by doubling the fractional uncertainty. For Sgr~A$^*$, the EHT image analysis, together with stellar-orbit mass-to-distance information, gives an inferred shadow-size deviation summarized in strong-gravity analyses as $\delta_{\rm Sgr A^*}\simeq -0.060\pm0.065$ \cite{EventHorizonTelescope:2022wkp,EventHorizonTelescope:2022urf,Vagnozzi:2022moj}. With the same conversion, we use

\begin{equation}
\textbf{{Sgr A*}} : 
\begin{cases}
4.55 \lesssim r_{\text{sh}}/M \lesssim 5.22 & (1\sigma) \\[6pt]
4.21 \lesssim r_{\text{sh}}/M \lesssim 5.56 & (2\sigma)
\end{cases}
\end{equation}

\begin{equation}
\textbf{M87*} : 
\begin{cases}
4.26 \lesssim r_{\text{sh}}/M \lesssim 6.03 & (1\sigma) \\[6pt]
3.38 \lesssim r_{\text{sh}}/M \lesssim 6.91 & (2\sigma)
\end{cases}
\end{equation}
Here $\sigma$ represents the confidence level of the effective shadow-size benchmark used in this geometric comparison.
We then translate the Sgr~A$^*$ and M$87^*$ effective shadow-size intervals above into the plasma-frequency ranges in Tables~\ref{tab:wc-ranges-sgr} and \ref{tab:wc-ranges-M87} as follows. For each source, plasma profile, and fixed set of black-hole/halo parameters, we compute the theoretical critical curve and evaluate
\begin{equation}
r_{sh}^{(P)}(x)=\sqrt{R_h^{(P)}(x)R_v^{(P)}(x)},\qquad x\equiv \frac{\omega_c^2}{\omega_0^2},
\end{equation}
where $P$ denotes the chosen plasma profile. The allowed interval for $x$ is the set of values for which the model radius lies inside the corresponding EHT-inferred benchmark interval,
\begin{equation}
x\in \mathcal{I}^{(P)}_{n\sigma}
\quad \Longleftrightarrow \quad
L_{n\sigma}\le r_{sh}^{(P)}(x)\le U_{n\sigma},
\end{equation}
with $(L_{n\sigma},U_{n\sigma})$ read from the corresponding Sgr~A$^*$ or M$87^*$ effective shadow-size interval. In practice, we sample $0<x<1$ and locate the endpoint of the interval by solving the appropriate crossing equation $r_{sh}^{(P)}(x)=L_{n\sigma}$ or $r_{sh}^{(P)}(x)=U_{n\sigma}$. For the homogeneous profile, $r_{sh}^{(P)}(x)$ increases over the plotted range, so the upper observational radius $U_{n\sigma}$ sets the quoted upper bound on $x$. For the inhomogeneous profile, the computed radius remains inside the listed benchmark interval over the full plotted range $0<x<1$, so the corresponding table entry is $0<x<1$. For the no-plasma case, there is no plasma-strength parameter to bound, so we mark the entry by ``--''.

\begin{table}[h!]
\centering
\begin{tabular}{|c|c|c|}
\hline
Plasma Profile & $1\sigma$ &  $2\sigma$ \\
\hline
No Plasma            & --      & -- \\

Inhomogeneous Plasma           & $0 < \frac{\omega_c^2}{\omega_0^2}  < 1$      & $0 < \frac{\omega_c^2}{\omega_0^2}  < 1$ \\
Homogeneous Plasma   & $0 < \frac{\omega_c^2}{\omega_0^2}  < 0.1123$ & $0 <\frac{\omega_c^2}{\omega_0^2}  <0.2593$ \\
\hline
\end{tabular}
\caption{\centering Illustrative geometric compatibility ranges of $\frac{\omega_c^2}{\omega_0^2}$ for different plasma profiles under the adopted Sgr~A$^*$ effective shadow-size interval.}
\label{tab:wc-ranges-sgr}
\end{table}

For $\text{Sgr A}^*$, the relevant parameter are $M_{BH} = 4.297 \times 10^6 M_{\odot}$ \cite{gravity_collaboration_polarimetry_2023}, $r_e= 12kpc, \rho_e =0.42 GeV cm^{-3} = 1.1 \times 10^7 M_{\odot}/kpc^3$ \cite{acevedo_dark_2025}, $\theta_0\approx\pi/2$, and $ a = 0.9$ \cite{sanchez_shadow_2024}. Table \ref{tab:wc-ranges-sgr} and Figure \ref{fig:sgr} show the geometric compatibility range of the plasma-strength parameter based on the $\text{Sgr A}^*$ effective shadow-size interval. For the inhomogeneous profile,  the shadow radius remains within the $2\sigma$ benchmark interval throughout the plotted interval $0<\omega_c^2/\omega_0^2<1$. For the homogeneous profile, larger values of $\omega_c^2/\omega_0^2$ inflate the geometric critical curve; under the adopted idealized mapping to the EHT-inferred size interval, values above 0.2593 would exceed the $2\sigma$ benchmark interval. \\
\begin{table}[h!]
\centering
\begin{tabular}{|c|c|c|}
\hline
Plasma Profile & $1\sigma$ &  $2\sigma$ \\
\hline
No Plasma            & --      & -- \\
Inhomogeneous          & $0 < \frac{\omega_c^2}{\omega_0^2}  < 1$      & $0 < \frac{\omega_c^2}{\omega_0^2}  < 1$ \\
Homogeneous Plasma   & $0 < \frac{\omega_c^2}{\omega_0^2}  < 0.369$ & $0 <\frac{\omega_c^2}{\omega_0^2}  < 0.559$ \\
\hline
\end{tabular}
\caption{\centering Illustrative geometric compatibility ranges of $\frac{\omega_c^2}{\omega_0^2}$ for different plasma profiles under the adopted M$87^*$ effective shadow-size interval.}
\label{tab:wc-ranges-M87}
\end{table}

For $\text{M87}^*$, the relevant parameter are $M_{BH} = 6.5 \times 10^9 M_{\odot}$, $r_e= 91.2kpc$, $\rho_e = 6.9 \times 10^6 M_{\odot}/kpc^3$ and $\theta_0=17^\circ$\cite{Liu:2023oab}. Table \ref{tab:wc-ranges-M87} and Figure \ref{fig:M87} show the geometric compatibility range of the plasma-strength parameter based on the M$87^*$ effective shadow-size interval. In the case of no plasma and inhomogeneous plasma, we observe that the geometric radius lies within the adopted $1\sigma$ benchmark interval throughout the plotted range. For the homogeneous profile, larger values of $\omega_c^2/\omega_0^2$ inflate the geometric critical curve; under the adopted idealized mapping, values above 0.369 would exceed the $1\sigma$ benchmark interval and values above 0.559 would exceed the $2\sigma$ benchmark interval.  

\section{Discussion and Future Work}\label{sect:ConclFW}
In this paper, we explored how the shadow of a Kerr-like black hole is influenced by the presence of plasma and a surrounding dark matter halo. We first analyzed the horizon structure of a black hole embedded in a dark matter background and found that, for sufficiently large dark matter densities, both the extremal spin and horizon radius increase, reflecting an overall enhancement of the gravitational potential. We discussed how the homogeneous plasma profile modifies the Bardeen celestial coordinates through the refractive normalization factor that arises from the leading-order coupling between $f_r(r)$ and $\Delta(r)$. Using Bardeen coordinates, we defined shadow observables (the shadow radius, deformation, and ellipticity) and compared our results with Event Horizon Telescope (EHT) measurements. As an illustrative benchmark, we found that the geometric critical-curve sizes can be compatible with the adopted EHT-inferred shadow-size intervals for both Sgr A* and M87* over the parameter ranges reported in Tables~\ref{tab:wc-ranges-sgr} and \ref{tab:wc-ranges-M87}. Although we modeled the dark matter halo using approximate parameters appropriate for each galaxy, their impact on the shadow observables is negligible. In contrast, our results show a significant difference between homogeneous and inhomogeneous plasma distributions, with increasingly distinct behavior emerging as the plasma strength is increased.

An interesting implication of these results is their relevance for interpreting EHT images of M87* and Sgr A* in the presence of plasma in the near-source, interstellar, and circumgalactic environments \cite{mathurCircumgalacticMediumMilky2015, misiriotisDistributionISMMilky2006}. Such material can modify photon propagation and may affect the relation between the mathematical critical curve and the observed brightness ring. The simplified plasma models employed here capture key qualitative features of dispersive propagation and may therefore serve as a useful starting point for more detailed ray-tracing studies. We showed only that the model can be geometrically compatible with EHT-inferred size intervals for certain values of the idealized plasma-strength parameter. A robust observational constraint would require constructing synthetic EHT images with the same refractive propagation model, emission prescription, scattering treatment, and image-analysis pipeline used to connect the critical curve to the observed brightness ring, which lies beyond the scope of the present study and provides a natural direction for future work.

Our present framework provides a controlled setting for exploring how plasma environments influence black hole shadow observables. The plasma models considered here are chosen to satisfy the common separability condition \cite{perlick_influence_2015}, which enables analytic treatment and facilitates clear interpretation of qualitative trends. Extending this analysis to more general plasma distributions, incorporating numerical ray tracing, would be a natural direction for future work.

\section*{Data Availability}
There are no publicly available research data or software supporting this manuscript. Requests for further information or data should be sent to the corresponding author. The observational benchmark values used in Sec.~\ref{sect:CEHTM87As} are taken from Refs.~\cite{EventHorizonTelescope:2021dqv,EventHorizonTelescope:2022wkp,EventHorizonTelescope:2022urf,Vagnozzi:2022moj}.

\begin{acknowledgments}
C. M. thanks the NSF LSAMP IINSPIRE Grant (NSF grant \#2207350, subgrant \#026450H) for funding this research. O. G., L. L., L.R., and S.R. acknowledge the support from the Grinnell College CSFS grant. We also thank the referee for the exceptionally careful reading of the manuscript and constructive suggestions, which greatly improved the paper's presentation and accuracy.
\end{acknowledgments}
\appendix
\bibliography{cftgr}

\end{document}